\documentclass[acmsmall,nonacm]{acmart}

\usepackage{multirow}
\usepackage{tcolorbox}
\usepackage{graphicx}
\usepackage{subcaption}
\usepackage{listings}
\usepackage{fancybox}

\usepackage{fancyhdr}
\AtBeginDocument{%
  }

\authorsaddresses{}

\fancypagestyle{acceptednotice}{
  \fancyhf{}
  \fancyhead[C]{\footnotesize \textcolor{gray}{This is the author's version of the work. Accepted to the 34th ACM International Conference on the Foundations of Software Engineering (FSE 2026), Montreal, Canada.}}
}

\fancypagestyle{acceptednotice}{
  \fancyhf{}
  \fancyhead[L]{\footnotesize \textsf{Accepted to FSE 2026}}
  \fancyhead[R]{\footnotesize \textsf{Author's Version}}
  \fancyfoot[C]{\footnotesize \itshape © Owner/Author 2026. This is the author's version of the work. It is posted here for your personal use. Not for redistribution.}
}

\begin{document}

\title[MetaRCA: A RCA Framework for Cloud-Native Systems Powered by Meta Causal Knowledge]{MetaRCA: A Generalizable Root Cause Analysis Framework for Cloud-Native Systems Powered by Meta Causal Knowledge}

\author{Shuai Liang}
\orcid{0009-0008-0256-9037}
\affiliation{%
  \institution{Sun Yat-sen University}
  \city{Guangzhou}
  \country{China}
}
\affiliation{%
  \institution{China Unicom Software Research Institute}
  \city{Beijing}
  \country{China}
}
\email{liangsh76@mail2.sysu.edu.cn}

\author{Pengfei Chen}
\orcid{0000-0003-0972-6900}
\affiliation{%
  \institution{Sun Yat-sen University}
  \city{Guangzhou}
  \country{China}
}
\email{chenpf7@mail.sysu.edu.cn}

\author{Bozhe Tian}
\orcid{0009-0005-6409-4506}
\affiliation{%
  \institution{China Unicom Software Research Institute}
  \city{Beijing}
  \country{China}
}
\email{tianbz11@chinaunicom.cn}

\author{Gou Tan}
\orcid{0009-0008-6580-1470}
\affiliation{%
  \institution{Sun Yat-sen University}
  \city{Guangzhou}
  \country{China}
}
\email{tang29@mail2.sysu.edu.cn}

\author{Maohong Xu}
\orcid{0009-0001-8208-6420}
\affiliation{%
  \institution{China Unicom Software Research Institute}
  \city{Beijing}
  \country{China}
}
\email{xumh6@chinaunicom.cn}

\author{Youjun Qu}
\orcid{0009-0008-9926-5899}
\affiliation{%
  \institution{China Unicom Software Research Institute}
  \city{Beijing}
  \country{China}
}
\email{quyj@chinaunicom.cn}

\author{Yahui Zhao}
\orcid{0009-0001-2024-4585}
\affiliation{%
  \institution{China Unicom Software Research Institute}
  \city{Beijing}
  \country{China}
}
\email{zhaoyh99@chinaunicom.cn}

\author{Yiduo Shang}
\orcid{0009-0005-5523-3084}
\affiliation{%
  \institution{China Unicom Software Research Institute}
  \city{Beijing}
  \country{China}
}
\email{shangyd3@chinaunicom.cn}

\author{Chongkang Tan}
\orcid{0009-0003-6822-6540}
\affiliation{%
  \institution{Individual Researcher}
  \city{Guangzhou}
  \country{China}
}
\email{chongk-t@hotmail.com}

\renewcommand{\shortauthors}{S. Liang, P. Chen, B. Tian, G. Tan, M. Xu, Y. Qu, Y. Zhao, Y. Shang, and C. Tan}


\begin{abstract}
  The dynamics and complexity of cloud-native systems present significant challenges for Root Cause Analysis (RCA). While causality-based RCA methods have shown significant progress in recent years, their practical adoption is fundamentally limited by three intertwined challenges: poor scalability against system complexity, brittle generalization across different system topologies, and inadequate integration of domain knowledge. These limitations create a vicious cycle, hindering the development of robust and efficient RCA solutions. This paper introduces \textit{MetaRCA}, a generalizable RCA framework for cloud-native systems. \textit{MetaRCA} first constructs a Meta Causal Graph (MCG) offline, a reusable knowledge base defined at the metadata level. To build the MCG, we propose an evidence-driven algorithm that systematically fuses knowledge from Large Language Models (LLMs), historical fault reports, and observability data. When a fault occurs, \textit{MetaRCA} performs a lightweight online inference by dynamically instantiating the MCG into a localized graph based on the current context, and then leverages real-time data to weight and prune causal links for precise root cause localization. Evaluated on 252 public and 59 production failures, \textit{MetaRCA} demonstrates state-of-the-art performance. It surpasses the strongest baseline by 29 percentage points in service-level and 48 percentage points in metric-level accuracy. This performance advantage widens as system complexity increases, with its overhead scaling near-linearly. Crucially, \textit{MetaRCA} shows robust cross-system generalization, maintaining over 80\% accuracy across diverse systems. 
\end{abstract}




\maketitle
\thispagestyle{acceptednotice}

% \footnote{This is the author-generated version based on the accepted manuscript. The final, official version of record will be available in \textbf{[FSE, 2026]}, published by \textbf{[Publisher: ACM]}.}

\section{Introduction}
Cloud-native architectures, built upon microservices, containerization, and dynamic orchestration, have become mainstream to modern digital infrastructure by enhancing system agility and elasticity\cite{zhang2024survey,howfar2024root,yao2024chain}. However, the very dynamism and distribution that provide these benefits also complicate fault propagation paths between services, making them difficult to diagnose. Consequently, performing fast and precise RCA has become a fundamental challenge to ensuring the reliability of these large-scale systems\cite{wang2021groot,pham2025rcaeval,li2022causal_circa}.

To address this challenge, automated RCA techniques have been a topic of active research and development in both academia and industry. Among these, causal inference-based methods\cite{wang2018cloudranger, meng2020localizing} have recently garnered significant attention due to their unique capability to reveal the logical relationships of fault propagation within a system. The core principle of this approach is to first analyze multi-dimensional time-series metrics to construct a causal graph that models the dependencies between services and metrics\cite{howfar2024root}. Subsequently, this graph is utilized to infer the root cause of a failure.

Early data-driven approaches for RCA constructed causal graphs by applying classical discovery algorithms(e.g., PC\cite{spirtes2000causation}) to system metrics. Methods like CloudRanger\cite{wang2018cloudranger} and AutoMap\cite{ma2020automap} then identify potential root causes on these graphs using ranking algorithms such as PageRank\cite{brin1998pagerank}. However, these purely data-driven techniques suffer from combinatorial explosion in large-scale systems. A subsequent line of work (e.g., CauseInfer\cite{chen2014causeinfer}, CIRCA\cite{li2022causal_circa}) prunes the search space by integrating the discovery process with the system's call dependency graph.

Furthermore, to fully leverage domain knowledge, some works\cite{liu2021microhecl,saha2022mining,wang2021groot,yao2024chain} first extract causal relationships from diverse sources like historical reports and expert knowledge, then construct these relationships into causal or event impact graphs, and finally employ graph algorithms to rank potential root causes.
Recent advancements in Large Language Models (LLMs) have spurred new approaches for structured knowledge extraction, with studies\cite{kiciman2023causal,xie2024cloudatlasefficientfault} demonstrating their potential in causal discovery, methods like RCACopilot\cite{chen2024automatic} and OpenRCA\cite{xu2025openrca} now directly apply the contextual reasoning of LLMs to perform root cause analysis.

Despite significant progress, the practical application of causal-based RCA is hindered by three challenges:
\textbf{(1) Scalability.} Current methods struggle to scale to the complexity of modern production systems\cite{howfar2024root}. Their computational cost, often growing super-exponentially with the number of components, prohibits the real-time analysis required for effective incident response.
\textbf{(2) Generalization.} The causal models generated are often brittle and overfitted to specific system topologies or historical data distributions. This lack of generalizability necessitates costly retraining for new system versions or instances, impeding their widespread adoption and reuse\cite{xie2024cloudatlasefficientfault}.
(3)\textbf{ Poor Knowledge Integration.} A fundamental dilemma exists in leveraging domain knowledge. Traditional approaches are confined to simplistic, pre-defined rules\cite{li2022causal_circa,liu2021microhecl} or system-specific knowledge that cannot be transferred\cite{wang2021groot,yao2024chain}.Conversely, while nascent LLM-based methods promise broader knowledge, they are plagued by hallucinations, high inference latency, and prohibitive operational costs, limiting their current utility\cite{kiciman2023causal}.
These challenges are deeply intertwined: the lack of knowledge fusion (Challenge 3) weakens the generalization ability of models (Challenge 2), which forces a reliance on large-scale, system-specific data, thereby exacerbating the scalability bottlenecks (Challenge 1).

To address these challenges, we propose \textbf{MetaRCA}, a generalizable RCA framework for cloud-native systems. The core insight of MetaRCA is to decouple the offline construction of a reusable, metadata-level causal knowledge base from the online, context-aware inference process. Offline, MetaRCA constructs a \textbf{Meta Causal Graph} (MCG) by systematically fusing multi-source knowledge from LLMs, historical incident reports, and monitoring data. At runtime, when a fault occurs, it dynamically instantiates the MCG into a localized, context-specific graph, prunes and weights it using real-time data, and finally applies a ranking method to pinpoint the root cause.

We conducted comprehensive experiments to validate MetaRCA's effectiveness. Our framework's core knowledge base the MCG, was constructed from a vast corpus comprising 563 production failure reports from China Unicom and 614 public failure datasets. 
We then evaluated MetaRCA's localization performance against five state-of-the-art baselines on 252 public cases from RCAEval and 59 real-world failures from China Unicom's production systems. The results are threefold: \textbf{(1) Accuracy}: MetaRCA significantly outperforms the strongest baseline, improving service-level and metric-level accuracy by 29 percentage points and 48 percentage points, respectively. \textbf{(2) Scalability}: This performance advantage widens as system complexity increases, while the computational overhead grows approximately linearly with system size. \textbf{(3) Generalization}: The framework demonstrates robust generalization, maintaining a stable accuracy of over 80\% across diverse systems without requiring retraining.

\textbf{Contributions.} In summary, this study makes the following contributions.
\begin{itemize}
    \item We propose MetaRCA, a generalizable framework for cloud-native RCA that systematically mitigates the critical challenges of scalability, generalization, and knowledge integration inherent in existing causality-based methods.
    \item We design the Meta Causal Graph, a reusable knowledge structure defined at the metadata level. built through an evidence-driven belief evolution algorithm that fuses heterogeneous knowledge from LLMs, incident reports, and monitoring data.
    \item We develop a lightweight online inference mechanism for fast and adaptive RCA, which at runtime instantiates, weights, and prunes a localized causal graph from the MCG using real-time data.
    \item We implemented MetaRCA and conducted a comprehensive evaluation on public benchmarks and a large-scale production system, demonstrating that MetaRCA achieves significant advantages in accuracy, scalability, and cross-system generalization.
\end{itemize}

\section{Background and Motivation}
\subsection{Background}
\begin{figure}[htbp]
    \centering 

    \begin{subfigure}{0.33\textwidth}
        \centering
        \includegraphics[width=\linewidth]{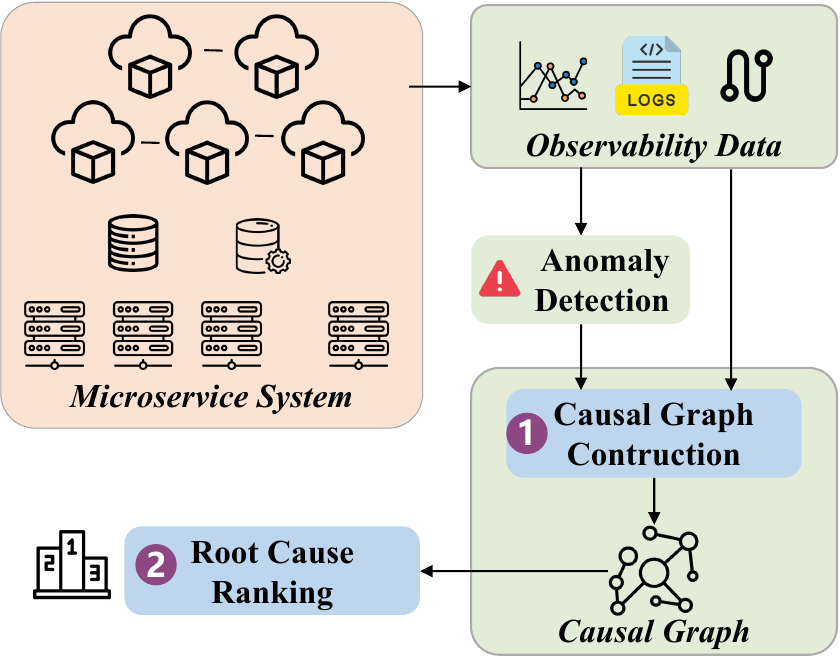}
        \caption{The causal RCA workflow.}
        \label{fig:causal-based rca}
    \end{subfigure}\hfill 
    \begin{subfigure}{0.6\textwidth}
        \centering
        \includegraphics[width=\linewidth]{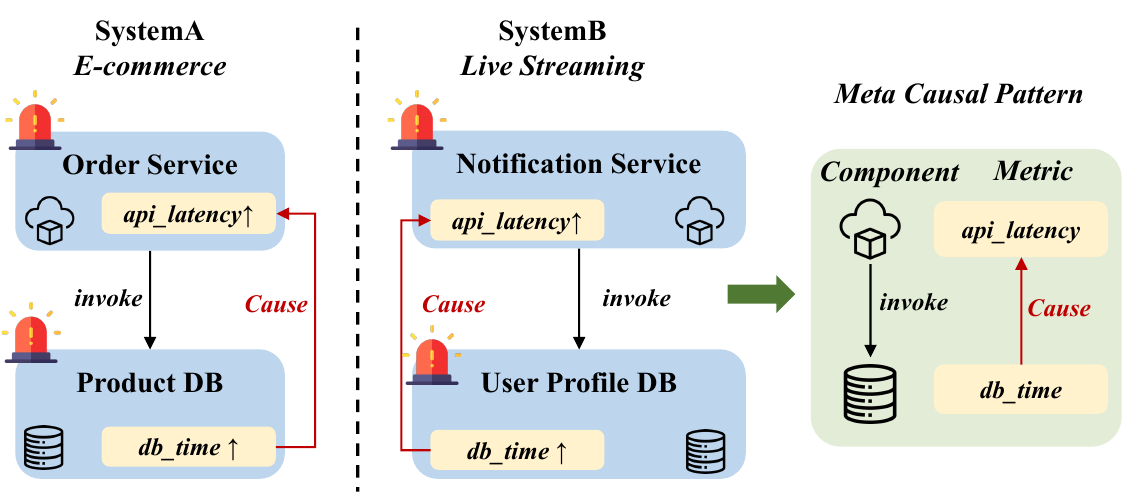}
        \caption{Cross-system causal pattern abstraction.}
        \label{fig:metacaual}
    \end{subfigure}

    \caption{Shifting from system-specific RCA to reusable causal patterns.}
    \label{fig:background}
\end{figure}

\subsubsection{\textbf{Root Cause Analysis based on Causal Graph}}
\label{sec:background_paradigm}
This paradigm conceptualizes fault localization as a structured, two-stage process, as illustrated in Figure~\ref{fig:background}(a).
The cornerstone is a mathematical model known as a \textit{Causal Graph}\cite{howfar2024root}, typically represented as a directed graph $G=(V, E)$. In the context of RCA, the vertex set $V$ represents observable system entities, such as service instances or fine-grained performance metrics (e.g., CPU utilization, API latency)\cite{howfar2024root}. Unlike traditional dependency graphs that merely describe static call relationships, the directed edge set $E$ is designed to capture the system's dynamic causal dependencies at runtime. For instance, in the e-commerce system example shown in Figure~\ref{fig:background}(b), an edge from the $db\_time$ metric of the Product DB to the $api\_latency$ metric of the Order Service signifies the causal relationship that "\textit{an increase in database processing time directly leads to an increase in the upstream service's interface latency.}"

Based on such a causal graph, the general RCA process proceeds as follows.
\begin{itemize}
\item \textit{Causal Graph Construction.} In the first stage, a localized causal graph is constructed to model how anomalies are propagating among relevant entities. This is achieved by applying causal discovery algorithms\cite{granger1980testing,spirtes2000causation} or leveraging domain knowledge on various observability data sources (e.g., metrics, logs, topology).
\item \textit{Root Cause Ranking.} In the second stage, with the causal graph obtained, a graph-ranking algorithm (e.g., PageRank\cite{brin1998pagerank}) is applied to perform inference. These algorithms analyze the graph's topology and edge weights to quantify the likelihood of each node being the "source" of the fault propagation, ultimately producing a ranked list of root cause candidates.
\end{itemize}

\subsubsection{\textbf{Fault Reports.}}
A fault report is a semi-structured knowledge document drafted by SRE teams after a failure has been resolved. These reports systematically summarize and retrospect a failure event, with a typical report encompassing several core elements.
\begin{itemize}
\item \textit{Fault symptoms and impact,} describing the specific anomalies perceived by users or the system, along with their business impact.
\item \textit{Fault handling process,} documenting the key steps taken during the emergency response and remediation.
\item \textit{Root cause analysis,} providing an in-depth analysis of the direct and fundamental causes that led to the failure.
\item \textit{Preventative and improvement measures,} proposing technical or procedural optimizations to prevent similar issues from recurring.
\end{itemize}
These documents encapsulate valuable expert experience and causal knowledge, thus representing a high-quality source of knowledge.

\subsection{Motivation}
Despite increasing system complexity, we identify three key observations that together underscore the feasibility and necessity of a generalizable RCA method.

\textbf{First, cloud-native systems are highly componentized.} A complex system is typically built from a finite set of standard component types (e.g., Microservices, MySQL, Redis) interacting via standardized interfaces\cite{xie2024cloudatlasefficientfault}. This modularity is crucial as it enables knowledge abstraction at the type level, rather than being confined to individual instances. Table~\ref{tab:Componentization} provides quantitative evidence of this principle across diverse open-source and production systems.

\begin{table}[h]
    \centering
    \caption{Empirical evidence of componentization in representative cloud-native systems.}
    \label{tab:Componentization}
    \begin{tabular}{l ccc ccc}
        \toprule
        & \multicolumn{3}{c}{\textbf{Open-Source Systems}} & \multicolumn{3}{c}{\textbf{Production Systems}} \\
        \cmidrule(lr){2-4} \cmidrule(lr){5-7}
        \textbf{Metric} & \textbf{OB} & \textbf{SS} & \textbf{TT} & \textbf{BS} & \textbf{CSS} & \textbf{APP} \\
        \midrule
        Service Scale                 & 12    & 15    & 64    & 4000+  & 2000+  & 1000+  \\
        Component Type Count          & 3     & 3     & 4     & 17    & 10    & 12    \\
        Component Connection Patterns & 3     & 3     & 4     & 19    & 11    & 12    \\
        \bottomrule
    \end{tabular}
    
    \vspace{1ex} % 
    \begin{minipage}{0.95\columnwidth} 
        \scriptsize % 
        \textit{Legend}: OB: Online Boutique; SS: Sock Shop; TT: Train Ticket; BS: Business System; CSS: Customer Service System; APP: Mobile APP. 
        \par \vspace{0.5ex} % 
        Data were sourced from \cite{howfar2024root} for open-source systems, and from internal architecture diagrams and CMDB for production systems. Production service scales are estimated due to confidentiality.
    \end{minipage}
\end{table}

\textbf{Second, the enterprise-wide push for technology stack unification}, a practice motivated by efficiency, cost, and risk management\cite{tech_google_Henderson_2017,tech_forbesStrategicAdvantage}, creates a fertile ground for shared, reusable causal patterns to emerge across formerly disparate systems.

\textbf{Finally, the troubleshooting expertise of SREs is highly transferable.} An experienced engineer can efficiently diagnose faults across different business systems because they possess generic causal knowledge that is independent of specific business logic. This is starkly illustrated by the two disparate incidents in Figure~\ref{fig:background}(b). In an e-commerce System A, high database processing time ($db\_time$) in a Product DB leads to high $api\_latency$ in its Order Service. In a separate live streaming System B, an analogous issue occurs where high $db\_time$ in a User Profile DB causes high latency in its Notification Service. An expert SRE would recognize that both incidents are manifestations of the same underlying, reusable \textit{Meta Causal Pattern}—that performance degradation in a database component directly impacts the latency of its upstream callers.

These observations and the limitations of existing methods lead us to propose the central idea that:\textbf{ by structuring and abstracting this transferable expert knowledge into a metadata-level causal knowledge base, and then leveraging it for reasoning over real-time system topology and monitoring data, we can achieve an automated RCA capability that "thinks" like an experienced SRE.}

\subsection{Problem Formulation}
Consider a system consisting of $N$ microservices, with the service set denoted by $\{s^i\}_{i=1}^N$. At each time step $t$, the monitoring system collects $M$ metrics from service $s^i$, denoted as $X_t^i=\{x_t^{(i,j)}\}_{j=1}^M$ ($M \ge 1$). Assume  a failure occurs at time $t_F$. The input to our method is a dataset $\mathbf{D}$ comprising all metrics from all services collected over the time window from a point $t_0$ before the failure to the time $t_{rca}$ when the RCA module is invoked, where $t_0 < t_F < t_{rca}$. The objective of a metric-based RCA method is to identify the root cause of the failure using the dataset $\mathbf{D}$.

\section{Detailed Design}

\begin{figure}
    \centering
    \includegraphics[width=0.95\linewidth]{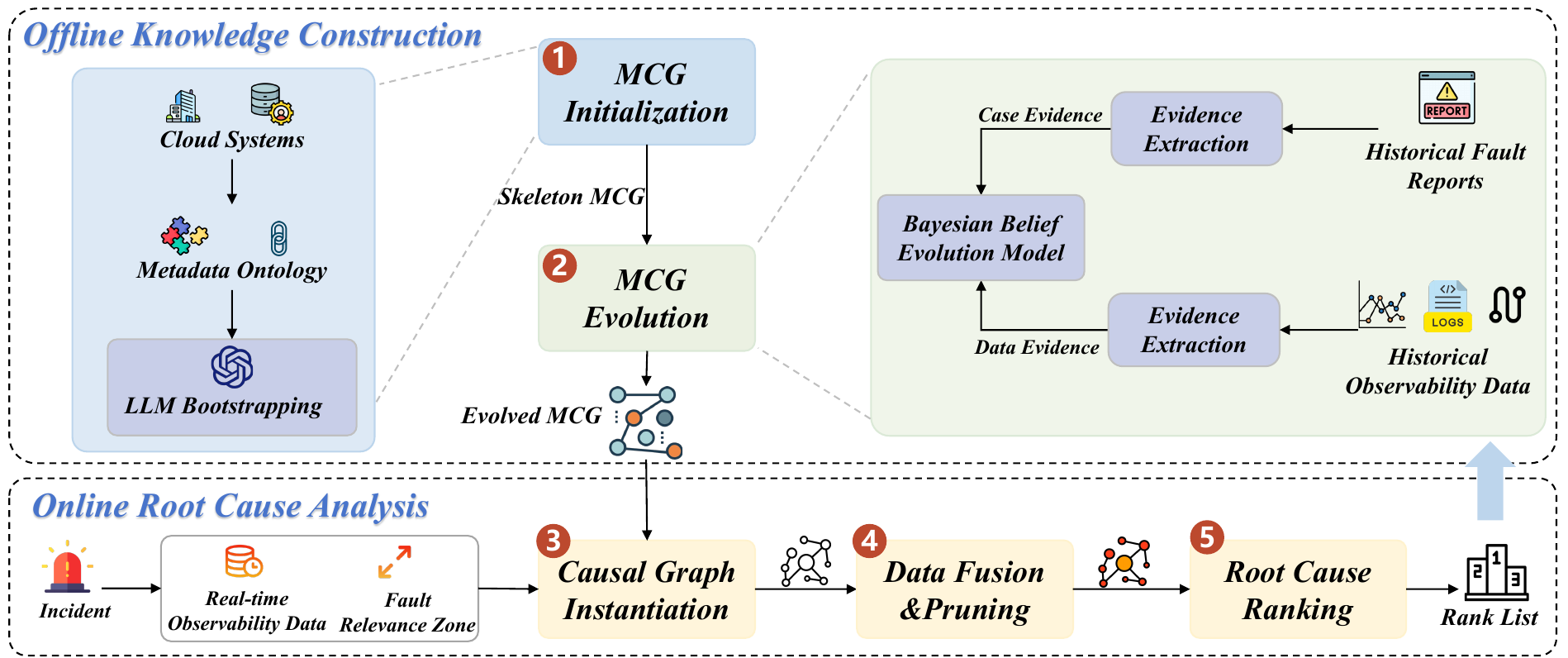}
    \caption{The framework of MetaRCA.}
    \label{fig:overview}
\end{figure}

This section details the design of MetaRCA, a generalizable RCA framework tailored for cloud-native systems. The overall architecture of our framework is depicted in Figure \ref{fig:overview}. MetaRCA operates in two distinct phases: an offline phase and an online phase. The offline phase consists of \textit{MCG Initialization} and \textit{MCG Evolution}, which are elaborated in \S\ref{sec:Initialization} and \S\ref{sec:evolution}, respectively. The online phase, \textit{Online Root Cause Analysis}, is detailed in \S\ref{sec:rca} and comprises three sequential sub-modules: \textit{Causal Graph Instantiation}, \textit{Data Fusion and Pruning}, and \textit{Root Cause Ranking}. 
\subsection{Meta Causal Graph Initialization}
\label{sec:Initialization}
The Meta Causal Graph (MCG) is a knowledge graph, describing the potential intra- and inter-component causal relationships at the metadata level, based on the types of components and their specific connection relationships.

\subsubsection{\textbf{Metadata Ontology Definition}}
\label{sec:metadata}

To formalize the MCG, we define the following metadata ontologies.
\begin{itemize}
    \item \textit{Component Types ($\mathcal{C}$)}: a set of all component types within the system, such as \{Microservice, MySQL, Redis, Host\}.
    \item \textit{Metrics ($\mathcal{M}$)}: for each component type $c \in \mathcal{C}$, a set $\mathcal{M}(c)$ defines its standardized metrics. Each metric is a tuple of its name and semantic meaning.
    \item \textit{Connection Patterns ($\mathcal{P}$)}: a set of all valid connection patterns between component types. Each pattern $p \in \mathcal{P}$ is a triple $(c_{src}, c_{dst}, conn\_type)$ defining a legal interaction template. Inspired by \cite{causalitymining}, we define two fundamental patterns:
    \begin{itemize}
        \item \textit{Invoke} represents a service invocation pattern, e.g., (Microservice, MySQL).
        \item \textit{On} represents a hosting or resource dependency pattern, e.g., (Microservice, Host).
    \end{itemize}
\end{itemize}

\subsubsection{\textbf{LLM-based Causal Knowledge Bootstrapping}}

We leverage LLMs to automatically extract causal knowledge, thereby bootstrapping the construction of an initial skeleton MCG.

Inspired by established principles for metric classification in monitoring systems\cite{2016Site,chen2014causeinfer,li2022causal_circa}, we categorize metrics into two distinct types based on their roles in service interactions:
\begin{itemize}
    \item \textit{Service level indicators (SLIs)}. These are metrics that quantify the quality of a service, such as latency and error rate. Their impact \textit{propagates directly} to upstream and downstream services along the dependency graph.
    \item \textit{Resource metrics}. These metrics measure the internal saturation of a component's resources, like CPU utilization or memory usage. They causally influence the SLIs from within the service boundary.
\end{itemize}

For each instance $p=(c_{src}, c_{dst}, conn\_type)$ in the set $\mathcal{P}$, we design a structured prompt to guide the LLM's reasoning process. The prompt instructs the LLM to infer all potential causal relationships and their directions between the metrics of $c_{src}$ (from $\mathcal{M}(c_{src})$) and $c_{dst}$ (from $\mathcal{M}(c_{dst})$), considering both SLIs and resource metrics within the context of a cloud-native system where $c_{src}$ interacts with $c_{dst}$.

Based on this process, a skeleton MCG is constructed. A causal edge $e$ in this graph, representing a relationship between a cause metric $m_i$ and an effect metric $m_j$, is formally defined as:
\begin{equation}
e(cause, effect) \triangleq (m_i, m_j)  \:   on \: (c_{src} \xrightarrow{conn\_type} c_{dst}), 
\end{equation}
where $m_i$ and $m_j$ can belong to either $\mathcal{M}(c_{src})$ or $\mathcal{M}(c_{dst})$, depending on the causal direction.

\subsection{Meta Causal Graph Evolution}
\label{sec:evolution}
The skeleton MCG obtained through LLM bootstrapping is a static, unverified theoretical model. To transform it into a robust knowledge base that incorporates real-world experience, quantifies confidence, and enables continuous evolution, we designed a multi-stage, evidence-driven belief evolution algorithm.

The algorithm first processes raw data via a standardized evidence extraction and alignment module. It then utilizes a belief update model equipped with time decay and perturbation suppression to continuously evolve the MCG.

\subsubsection{\textbf{Evidence Extraction}}
\label{sec:evidence}
To ensure that evidence from various sources can be reliably integrated into the MCG, we designed a standardized process for transforming raw information into structured evidence units.

\textbf{Extracting case evidence from failure reports ($\mathcal{E}_{fr}$).}
Historical failure reports, confirmed by human experts, are a high-quality source of causal knowledge. To systematically extract this knowledge, we employ a two-stage prompting strategy that leverages the in-context reasoning capabilities of Large Language Models: 

\textit{Causal entity extraction}, instructs an LLM to identify component and metric entities and their causal relationships from the "failure symptom" and "root cause analysis" sections of fault report.

\textit{Metric semantic alignment}, a constrained prompt instructs the LLM to align the extracted entities with the standard component types and metric names defined in our metadata ontology (\S\ref{sec:metadata}).

The final output of this process is a structured set of case evidence $e_{fr}$. Each evidence is formatted as follows:
\begin{equation}
e_{fr} = \{(c_{cause}, m_{cause}), (c_{effect}, m_{effect}), t_{report}\},
\end{equation}
where $(c_{cause}, m_{cause})$ and $c_{effect}, m_{effect}$ represent the cause entity tuple and effect entity tuple, respectively, and $t_{report}$ is the timestamp of the failure report.
    
\textbf{Extracting statistical evidence from historical observability data ($\mathcal{E}_{da}$).}
We derive objective statistical evidence from historical data through a two-stage refinement process:

\textit{Causal discovery.}
For each failure incident, we begin by extracting the relevant multivariate metric time series. We then apply a causal discovery algorithm (e.g., PC\cite{spirtes2000causation}, PCMCI~\cite{pcmci_2019}) to the time series data. The output of this stage is a set of raw, instance-level causal links that are statistically significant (i.e., pass the $p < \alpha$ test). These links represent potential causal relationships between specific metrics of service instances.

\textit{Connection patterns alignment.}
We first prune causal links that are inconsistent with the system's service dependency graph, retaining only those that align with known upstream/downstream relationships. The resulting instance-level links are then abstracted into meta-level relationships by aligning them with predefined \textit{Connection Patterns} (\S\ref{sec:metadata}). For example, this transforms an instance-level link like \textit{(service\_A.latency, service\_B.latency)} into the meta-level \textit{(microservice.latency, microservice\_another.latency)}. Each meta-level pair is formulated into a standardized statistical evidence unit, $e_{da}$:
\begin{equation}
    e_{da} = \{(c_{\text{cause}}, m_{\text{cause}}), (c_{\text{effect}}, m_{\text{effect}}), t_{\text{data}}\}, 
\end{equation}
where $(c_{\text{cause}}, m_{\text{cause}})$ and $(c_{\text{effect}}, m_{\text{effect}})$ represent the meta-level cause and effect, and $t_{\text{data}}$ is the timestamp of fault time. 
All such units form the statistical evidence set $\mathcal{E}_{da}$.

\subsubsection{\textbf{Bayesian Belief Evolution Model}}

Having obtained standardized evidence sets, we employ a Bayesian belief evolution model to dynamically update the \textit{Causal Belief Score (CBS)} for each edge in the MCG. This model operates in the \textit{Log-Odds} space for numerical stability, computational efficiency, and elegance, representing a sequential application of Bayes' theorem\cite{kessler2023sequential}.

\textbf{Model definition.}
The belief score $CBS(e)$ for any edge $e$, representing the posterior probability $P(H_e | \mathcal{E})$ of the causal hypothesis $H_e$ given all evidence $\mathcal{E}$, is determined by an underlying Aggregated \textit{Log-Odds Score} $L(e)$. The conversion is performed via the standard Logistic (Sigmoid) function:
\begin{equation}
CBS(e) = \sigma(L(e)) = \frac{1}{1 + \exp(-L(e))}.
\end{equation}

The process begins with a prior belief bootstrapped from the LLM, denoted as $p_0(e)$. This prior is converted into an initial \textit{Log-Odds} score $L_0(e)$, which serves as the starting point for all subsequent Bayesian updates, representing a sequential application of Bayes' theorem
\begin{equation}
L_0(e) = \log\left(\frac{p_0(e)}{1 - p_0(e)}\right).
\end{equation}

We can assume the belief score given by the LLM is $p_0(e)=0.5$, which results in an initial $L_0(e)=0$.

\textbf{Offline batch weighting.} 
During initial system deployment or periodic recalibration, we compute the posterior belief by incorporating all available historical evidence in a batch process. For each edge $e$, its posterior score $L_{\text{batch}}(e)$ is calculated by updating the prior $L_0(e)$ with the sum of all evidence contributions:
\begin{equation}
L_{\text{batch}}(e) = L_0(e) + \sum_{fr_i \in \mathcal{E}_{fr}(e)} \Delta L_{fr}(fr_i) + \sum_{da_j \in \mathcal{E}_{da}(e)} \Delta L_{da}(da_j),
\end{equation}
where $\mathcal{E}_{fr}(e)$ and $\mathcal{E}_{da}(e)$ are the evidence sets for edge $e$. Each increment $\Delta L$ represents the \textit{Log-Bayes Factor} of a piece of evidence, quantifying its supporting strength. It is calculated as follows, incorporating evidence type asymmetry and a temporal decay function, $\text{Decay}(\Delta t)$:

\begin{equation}
\Delta L_{fr} = \lambda_{fr} \cdot \text{Decay}(\Delta t)
\qquad
\Delta L_{da} = \lambda_{da} \cdot \text{Decay}(\Delta t).
\end{equation}

Here, $\lambda_{fr}$ and $\lambda_{da}$ are the base \textit{Log-Bayes Factors} for high-confidence (fault report) and low-confidence (data-driven) evidence, respectively, with $\lambda_{fr} \gg \lambda_{da}$. The decay function is defined as $\text{Decay}(\Delta t) = \exp(-k \cdot \Delta t)$, where $\Delta t$ is the age of the evidence and $k$ is the decay constant.

This batch process also evolves the MCG's structure. To maintain graph integrity, only high-confidence case evidence can introduce new causal edges. When such evidence for a non-existent edge is found, the edge is created with a neutral prior ($L_0=0$), and its $L_{\text{batch}}$ score is then computed using all relevant historical evidence for it.

\textbf{Online streaming update.} 
Following the batch initialization, the MCG undergoes continuous evolution via sequential Bayesian updating. When a new evidence arrives, the system performs a lightweight streaming update. This two-step process first accounts for the passage of time (belief decay) and then incorporates the new evidence (belief augmentation).

First, the current \textit{Log-Odds} score is decayed to reflect that older beliefs are less certain. The decayed score serves as the new transient prior:
\begin{equation}
L_{\text{prior}'}(e) = L_{\text{current}}(e) \cdot \exp(-k \cdot (t_{\text{current}} - t_{\text{last\_update}}(e))).
\end{equation}
Next, this transient prior is updated with the Log-Bayes Factor of the new evidence to yield the new posterior score:
\begin{equation}
L_{\text{new}}(e) = L_{\text{prior}'}(e) + \Delta L_{\text{new}}.
\end{equation}
The increment $\Delta L_{\text{new}}$ is computed based on the evidence type, using the same base impact strengths defined in the batch process. This hybrid approach, combining robust batch initialization with agile online updates, ensures the MCG's long-term adaptability to system evolution.

\subsection{MCG-based Online Root Cause Analysis}
\label{sec:rca}

With the offline MCG as a foundation, our online analysis pipeline first instantiates a localized causal graph from observed symptoms (\textit{Causal Graph Instantiation}), then refines and prunes it by fusing real-time data (\textit{Data Fusion and Pruning}), and finally performs inference to rank the most probable root causes (\textit{Root Cause Ranking}).

\subsubsection{\textbf{Causal Graph Instantiation}}
The objective of this phase is to construct an instance-level "skeleton" causal graph relevant to the current fault context from the monitoring data.

\textbf{Fault relevance zone delimitation.}
We first apply a time-series anomaly detection algorithm (e.g., $3\sigma$ with baseline) to identify all anomalous metric instances. The set of components exhibiting at least one such metric constitutes the Fault Relevance Zone (FRZ).

\textbf{Instantiation.}
Subsequently, a \textit{Localized Instance Causal Graph (LICG)} is instantiated by projecting meta-causal relationships from the MCG onto component connections within the FRZ. Each edge inherits its \textit{Causal Belief Score} from the MCG as \textit{a priori} knowledge.

\subsubsection{\textbf{Data Fusion and Pruning.}}
This phase refines the instantiated LICG by confronting it with real-time observability data. The process fuses the static  priori belief with a score derived from current observations, then prunes the graph to produce a high-fidelity causal graph for inference.

\textbf{Contextual plausibility score calculation.}
For each edge $e_i: u \to v$, we compute a \textit{Contextual Plausibility Score} $S_{\text{context}}(e_i) \in [0, 1]$. This score is the product of two observational factors:
\begin{equation}
S_{\text{context}}(e_i) = S_{\text{anomaly}}(u, v) \times S_{\text{corr}}(u, v).
\end{equation}
The first factor, the \textit{Anomaly Co-occurrence Score ($S_{\text{anomaly}}$)}, verifies that both nodes exhibit significant anomalous behavior. It is computed by applying a logistic function to the maximum absolute z-scores of both time series and combining the results with a $min$ operator:
\begin{equation}
S_{\text{anomaly}}(u, v) = \min\left(\sigma(|Z_u|_{\max} ), \sigma(|Z_v|_{\max} )\right),
\end{equation}
where $\sigma(\cdot)$ is the sigmoid function. The second factor, the \textit{Lagged Correlation Score ($S_{\text{corr}}$)}, robustly quantifies both pattern similarity and temporal precedence. It is calculated as the maximum \textit{Pearson Correlation} between the cause time series ($\mathcal{T}_u$) and the effect time series ($\mathcal{T}_v$) over a set of small, non-negative lags ($k$):
\begin{equation}
S_{\text{corr}}(u, v) = \max_{k \in {\{0, 1, ..., k_{\max}\}}} \left| \rho(\mathcal{T}_u(t), \mathcal{T}_{v}(t-k)) \right|,
\end{equation}
where $\rho(\cdot, \cdot)$ denotes the standard \textit{Pearson Correlation Coefficient} \cite{pearson1895vii}, where $\mathcal{T}_{v}(t-k)$ is the time series of node $v$ lagged by $k$ steps at time $t$.

\textbf{Causal weight fusion and pruning.}
Finally, the \textit{Causal Weight} $W_{LICG}(e_i)$ is computed by fusing the a priori belief $W_{MCG}(e_i)$ with the contextual score:
\begin{equation}
W_{LICG}(e_i) = W_{MCG}(e_i) \times S_{\text{context}}(e_i).
\end{equation}
Edges where $W_{LICG}(e_i)$ falls below a threshold $\theta_p$ are then pruned. We set this threshold to 0.3 by default and will provide a detailed description in \S\ref{sec:discussion}.

\subsubsection{\textbf{Root Cause Ranking}}
\label{sec:ccb}
After obtaining the refined LICG, $G_L=(V_L, E_L)$, we employ an iterative algorithm, termed \textit{Causal Contribution Back-propagation (CCB)}, for the final root cause ranking. The algorithm is designed to compute an importance score for each node by aggregating its intrinsic anomaly score with the weighted fault contributions back-propagated from its downstream effects.

Formally, the algorithm computes a fault contribution score for each node $u \in V_L$. Starting with a score of 1 for all nodes, the score is updated in each iteration $k+1$ as follows:
\begin{equation}
\text{Score}_{k+1}(u) =  \sum_{v \in \text{Children}(u)} \left[ \text{Score}_k(v) \cdot W_{LICG}(e_{u,v}) \right].
\end{equation}
The iteration continues until the scores converge($Score_{k+1}-Score_k<\epsilon$), and the final scores are used to rank all nodes in descending order, yielding the $top-K$ candidate root causes.

\section{Experiment Evaluation}

We conduct a comprehensive evaluation of the MetaRCA framework on a diverse set of datasets, including three public microservice benchmarks and a real-world production dataset. Our evaluation is designed to systematically answer the following four core research questions (RQs).

\begin{itemize}
\item \textbf{RQ1: Effectiveness.} How does MetaRCA's localization accuracy and efficiency perform compared to baseline methods?

\item \textbf{RQ2: Generalizability and Scalability.} How effectively does MetaRCA generalize to diverse systems and scale with increasing system complexity?

\item \textbf{RQ3: Causal Graph Quality.} Is the causal graph generated by our approach, which fuses meta-causal knowledge and data, superior to those constructed by existing methods?

\item \textbf{RQ4: Ablation Study.} How do the key design components of MetaRCA individually contribute to its final localization performance?
\end{itemize}

\subsection{Implementation of MCG}
\label{sec:mcg_implementation}

To ensure the reproducibility of our experiments, this section details the specific implementation and configuration process of MCG.

\subsubsection{\textbf{Skeleton MCG Implementation}}

\textbf{Metadata ontology.} The foundation of our methodology is a predefined metadata ontology covering mainstream cloud-native technologies. This ontology provides a unified semantic basis for all subsequent knowledge extraction and alignment. In our experiments, we developed an ontology comprising \textbf{17} predominant component types and their valid connection patterns, derived from a comprehensive study of 36 systems (32 production, 4 open-source). These architectural elements were systematically extracted from CMDBs and deployment diagrams, then further refined through expert interviews to ensure practical fidelity. The metric set for each component follows a comprehensive, standardized list based on enterprise-level best practices from {China Unicom}, with its SLIs annotated by domain experts.

\textbf{LLM-based knowledge bootstrapping.} Building upon the metadata ontology, we utilized \textit{Google Gemini 2.5 Flash}\cite{gemini} to bootstrap an initial skeleton MCG. Appendix \ref{sec:prompt} provides the detailed prompt templates. The initial CBS for all bootstrapped edges was set to 0.5.

\subsubsection{\textbf{Implementation of Case Evidence Extraction}}

We extracted case evidence from a large and diverse corpus of incident reports. This corpus consists of \textbf{563} fault reports from over 100 real-world production systems within {China Unicom}, spanning a period from January 2020 to March 2025. These reports, having undergone multiple internal cycles of writing, auditing, and review, are of high quality and credibility. They cover a wide range of business domains (e.g., service subscription, customer support, mobile applications) and fault types across the entire technology stack, from infrastructure to the application layer.

We employed the two-stage process of \textit{Entity and Relation Extraction} followed by \textit{Semantic Alignment}, as described in \S\ref{sec:evidence}, to extract causal knowledge from these unstructured reports. Specifically, we used a privately deployed \textit{DeepSeek R1-70B}\cite{githubGitHubDeepseekaiDeepSeekR1} model for the information extraction task, with detailed prompts provided in Appendix \ref{sec:prompt}. 

\subsubsection{\textbf{Implementation of Statistical Evidence Extraction}}
To augment the MCG with statistical evidence, we utilized two public datasets: 375 fault cases from RCAEval-RE1\cite{pham2025rcaeval} and 239 fault cases from the AIOPS2022\cite{aiops22} challenge (cloudbed1, from March 28th to April 2nd).

For each fault case, we applied PCMCI \cite{pcmci_2019} as the causal discovery algorithm with a significance level $\alpha$ set to 0.05, due to its ability to handle high-dimensional time-series data. We strictly followed the two-stage refinement process described in \S\ref{sec:evidence} to carry out the implementation.

\subsubsection{\textbf{Evidence Fusion Configuration}}

The core hyperparameters in our evidence fusion and belief evolution algorithm were set as follows, we will discuss this further in \S\ref{sec:discussion}:
\begin{itemize}
    \item The base impact strength is set to 0.5 for incident reports($\lambda_{fr}$) and 0.05 for data-driven evidence($\lambda_{da}$).
    \item Temporal decay constant, $k = 0.005$ (with decay calculated in days),
\end{itemize}
Following this entire procedure, we generated the final MCG used for all subsequent evaluations.

\subsection{Datasets}
\label{sec:datasets}

Our evaluation is conducted on two distinct categories of datasets to validate the effectiveness of MetaRCA in both industrial settings and controlled academic environments: a dataset from a real-world, large-scale production environment and a widely-used public benchmark for RCA research. The details are summarized in Table~\ref{tab:eval_datasets}.

\begin{table}[h!]
  \caption{Overview of the evaluation datasets.}
  \label{tab:eval_datasets}
  \begin{tabular}{@{}llcc@{}}
    \toprule
    \textbf{Datasets} & \textbf{System Benchmark} & \textbf{\# of Cases} & \textbf{\# of Services} \\
    \midrule
    \textbf{\textit{Production}} & 4 production systems & 59 & 9 -- 112 \\
    \midrule
    \textbf{\textit{RE2-OB}} & Online Boutique & 90 & 12 \\
    \textbf{\textit{RE2-SS}} & Sock Shop & 90 & 15 \\
    \textbf{\textit{RE2-TT}}& Train Ticket & 72\textsuperscript{*} & 64 \\
    \bottomrule
  \end{tabular}
  \par
  \small 
  \raggedright
  \textsuperscript{*} 18 cases related to the ts-auth service were excluded because its absence from trace data made its call relationships untraceable.
  \par
\end{table}

\textbf{Production dataset setup.}
This dataset was collected from two mission-critical, large-scale production systems at China Unicom: the cBSS \footnote{cBSS is one of the world's most centralized Business Support Systems (BSS) with the largest single-instance user base in the telecom industry.} and the customer service system. Both systems serve over 400 million users and comprise thousands of microservices. To focus our study and ensure manageable analysis scopes, we treat each major business module within these systems as an independent system. We collected \textbf{59} real fault incidents that occurred between June and August 2025 from four representative subsystems. These incidents, while limited in number, are of high fidelity and complexity, spanning a wide range of system scales (from 9 to 112 services) and encompassing diverse, multi-level fault types, such as microservice CPU overload, third-party service latency, database congestion, and traffic surges. Crucially, these 59 incident cases are distinct and do not overlap with the incident reports used for the offline MCG construction, ensuring a fair evaluation on unseen faults. For each case, the ground truth root cause service and metric were meticulously annotated by SREs and technical experts.

\textbf{Open-source dataset setup.}
To compare our approach with prior work, we utilize the second dataset provided by \textbf{RCAEval} \cite{pham2025rcaeval}. This dataset contains \textbf{252} fault cases generated from three open-source microservice systems. The faults, covering 6 different types, were systematically injected, and the corresponding multi-variate observability data before and after the injection points were collected. The injection point serves as the ground truth for each case.

\subsection{Baseline Methods}
\label{sec:baselines}

For a comprehensive evaluation, we compare MetaRCA against five representative, causality-based baselines: two top-performing methods identified in the survey by\cite{howfar2024root}, two causal discovery-based approaches, and one recent method leveraging LLM.

\begin{itemize}
    
    \item \textbf{CIRCA}\cite{li2022causal_circa},a knowledge-driven approach, which constructs a causal architectural graph based on predefined causal assumptions and then uses regression-based hypothesis testing for inference.
    \item \textbf{CausalRCA}\cite{xin2023causalrca}, employs a gradient-based method to learn a weighted causal graph from data, followed by the PageRank algorithm for ranking.
    \item \textbf{PC\_PR}, the classical statistical approach, which first constructs a causal graph from metrics using the PC algorithm \cite{spirtes2000causation} and then applies PageRank\cite{howfar2024root} for root cause scoring.
    \item \textbf{PCMCI\_RW},which first constructs a causal graph from time-series metrics using the PCMCI\cite{pcmci_2019} algorithm and then applies a random walk\cite{howfar2024root} for root cause scoring.
    \item \textbf{OpenRCA}\cite{xu2025openrca}, leverages a large language model as the knowledge brain to drive a python executor for data analysis, iteratively analyzing results to guide subsequent actions and achieve precise root cause identification of faults.
\end{itemize}

\subsection{Evaluation Metrics}
\label{sec:metrics}

To assess the performance of MetaRCA and the baseline methods, we employ a set of standard metrics that evaluate both the accuracy and efficiency of the RCA task. Following existing works \cite{li2022causal_circa,howfar2024root,chen2014causeinfer}, we conduct evaluations at two granularities: the coarse-grained service level and the fine-grained metric level.

\textit{Accuracy.}
We use AC@k to measure the probability that the true root cause is included in the top-k ranked results provided by a method. Given a set of failure cases A, AC@k is calculated as:
\begin{equation}
\text{AC@k} = \frac{1}{|\mathcal{A}|} \sum_{a \in \mathcal{A}} \mathbb{I}(\text{rank}(g_a) \le k),
\end{equation}
where $g_a$is the ground truth root cause for failure case a, rank($g_a$) is its rank in the output list, and $\mathbb{I}(\cdot)$ is the indicator function. We report AC@k for both the root cause service (AC@k\_service) and the more precise root cause metric (AC@k\_metric), with k values of 1, 3, and 5.

\textit{Efficiency.}
We measure the \textit{Average RCA Time} ($\bar{T}_{rca}$) to evaluate the efficiency of each method. For a set of failure cases $\mathcal{A}$, this is defined as the average end-to-end wall-clock time (in seconds) from the moment an anomaly is first detected ($t_{detect}$) to the moment the algorithm outputs the final ranked list of root cause candidates ($t_{output}$):
\begin{equation}
\bar{T}_{rca} = \frac{1}{|\mathcal{A}|} \sum_{a \in \mathcal{A}} (t_{output}(a) - t_{detect}(a)).
\end{equation}

We conduct all experiments on Linux servers equipped with 64 CPUs, 128GB RAM.

\subsection{RQ1: Performance of MetaRCA}
To evaluate the overall effectiveness of MetaRCA, we conducted an end-to-end comparison against five baseline methods on all datasets. The results, summarized in Table~\ref{tab:end_to_end_performance}, demonstrate the competitive performance and unique advantages of our framework.

\textit{Accuracy.}
As shown in Table~\ref{tab:end_to_end_performance}, MetaRCA demonstrates a consistently strong and often superior performance, especially in the most complex and realistic scenarios. In the real-world \textit{Production} dataset, MetaRCA achieves a Top-1 service-level accuracy (AC@1\_service) of \textbf{0.66}, which is a \textbf{73.7\%} improvement over the next best baseline (CIRCA at 0.38). This significant accuracy gain is even more pronounced at the fine-grained metric level, where our method achieves an AC@1\_metric of \textbf{0.54}, more than doubling the performance of the best alternative. A similar robust performance is observed in the RE2-TT dataset, which features a larger number of services.

On the RE2-SS dataset, while CIRCA achieves the highest Top-1 service accuracy, MetaRCA remains highly competitive (0.79 vs. 0.86) and demonstrates superior performance at the more challenging metric level (e.g., an AC@5\_metric of 0.66). This highlights MetaRCA's balanced and consistent performance across diverse environments, as it excels at both granularities in the more complex RE2-TT dataset, whereas CIRCA's metric-level accuracy drops significantly.

Interestingly, the OpenRCA method performs better on the \textit{Production} dataset than on the other three. We believe this is because the realism of the production cases is better suited to its knowledge-based logical reasoning.

\textit{Efficiency.}
The last column of Table~\ref{tab:end_to_end_performance} presents the average RCA time ($\bar{T}_{rca}$(s)). Due to its effective fusion of a knowledge base with localized data analysis, MetaRCA achieves excellent efficiency across all datasets while delivering high accuracy. For instance, its RCA time in the \textit{Production} dataset (0.9s) is comparable to CIRCA's (0.8s), but it is orders of magnitude faster than the learning-based CausalRCA (281s) and the LLM-based OpenRCA (384s). This demonstrates that our approach successfully avoids the high computational overhead typical of complex learning or reasoning models.

% \begin{tcolorbox}

% \textbf{Answering RQ1:} \textit{MetaRCA} achieves a robust, state-of-the-art accuracy, particularly in the complex, real-world \textit{Production} dataset. It provides the best overall balance of high accuracy across both service and metric levels, and crucially, combines this with exceptional efficiency, positioning it as a practical solution for production systems.
% \end{tcolorbox}

\begin{center}
    \cornersize{0.15} 
    \setlength{\fboxsep}{8pt} 
    \Ovalbox{
        \begin{minipage}{0.92\columnwidth}
            \textbf{Answering RQ1:} \textit{MetaRCA} achieves a robust, state-of-the-art accuracy, particularly in the complex, real-world \textit{Production} dataset. It provides the best overall balance of high accuracy across both service and metric levels, and crucially, combines this with exceptional efficiency, positioning it as a practical solution for production systems.
        \end{minipage}
    }
\end{center}

\begin{table*}[!h]
  \caption{Performance comparison of MetaRCA and baseline methods across all datasets. The values in parentheses represent the standard deviation. }
  \label{tab:end_to_end_performance}
  \resizebox{0.9\textwidth}{!}{% 
  \begin{tabular}{@{}llccccccc@{}}
    \toprule
    \multirow{2}{*}{\textbf{Dataset}} & \multirow{2}{*}{\textbf{Method}} & \multicolumn{2}{c}{\textbf{AC@1}} & \multicolumn{2}{c}{\textbf{AC@3}} & \multicolumn{2}{c}{\textbf{AC@5}} & \multirow{2}{*}{\textbf{$\bar{T}_{rca}$(s)}} \\
    \cmidrule(lr){3-4} \cmidrule(lr){5-6} \cmidrule(lr){7-8}
    & & service & metric & service & metric & service & metric & \\
    \midrule
    \multirow{6}{*}{RE2-OB} 
    & PC\_PR & 0.17 & 0.06 & 0.29 & 0.11 & 0.52 & 0.17 & 4.07(0.26) \\
    & CausalRCA & 0.20 & 0.04 & 0.42 & 0.10 & 0.57 & 0.14 & 74.90(9.7) \\
    & PCMCI\_RW & 0.31 & 0.10 & 0.62 & 0.30 & 0.71 & 0.37 & 14.24(1.07) \\
    & CIRCA & 0.27 & 0.17 & 0.79 & 0.42 & \textbf{0.92} & \textbf{0.60} & 0.07(0.01) \\
    & OpenRCA & 0.23 & 0.00 & / & / & / & / & 222(85.5) \\
    & MetaRCA & \textbf{0.66} & \textbf{0.21} & \textbf{0.90} & \textbf{0.52} & 0.91 & \textbf{0.60} & 0.05(0.01) \\
    \midrule
    \multirow{6}{*}{RE2-SS} 
    & PC\_PR & 0.20 & 0.02 & 0.40 & 0.13 & 0.53 & 0.18 & 1.73(0.41) \\
    & CausalRCA & 0.17 & 0.04 & 0.31 & 0.07 & 0.51 & 0.13 & 60.1(6.53) \\
    & PCMCI\_RW & 0.16 & 0.01 & 0.33 & 0.03 & 0.54 & 0.04 & 18.47(1.05) \\
    & CIRCA & \textbf{0.86} & \textbf{0.29} & \textbf{0.91} & \textbf{0.60} & \textbf{0.94} & 0.62 & 0.04(0.01) \\
    & OpenRCA & 0.13 & 0.00 & / & / & / & / & 189.59(37.6) \\
    & MetaRCA & 0.79 & 0.19 & 0.89 & 0.59 & 0.90 & \textbf{0.66} & 0.05(0.01) \\
    \midrule
    \multirow{6}{*}{RE2-TT} 
    & PC\_PR & 0.03 & 0.01 & 0.08 & 0.04 & 0.13 & 0.06 & 4142.63(1735) \\
    & CausalRCA & 0.04 & 0.01 & 0.14 & 0.03 & 0.21 & 0.07 & 520.81(78.97) \\
    & PCMCI\_RW & 0.05 & 0.00 & 0.19 & 0.03 & 0.22 & 0.03 & 596.70(73.5) \\
    & CIRCA & 0.64 & 0.17 & 0.72 & 0.17 & 0.78 & 0.17 & 0.19(0.04) \\
    & OpenRCA & 0.1 & 0.03 & / & / & / & / & 290.23(128.6) \\
    & MetaRCA & \textbf{0.75} & \textbf{0.33} & \textbf{0.88} & \textbf{0.76} & \textbf{0.88} & \textbf{0.86} & 0.19(0.03) \\
    \midrule
    \multirow{6}{*}{Production} 
    & PC\_PR & 0.05 & 0.02 & 0.15 & 0.03 & 0.20 & 0.03 & 90.87 \\
    & CausalRCA & 0.10 & 0.03 & 0.24 & 0.05 & 0.32 & 0.08 & 281.82 \\
    & PCMCI\_RW & 0.14 & 0.05 & 0.22 & 0.16 & 0.28 & 0.16 & 338.25 \\
    & CIRCA & 0.38 & 0.12 & 0.59 & 0.34 & 0.76 & 0.44 & 0.80 \\
    & OpenRCA & 0.2 & 0.1 & / & / & / & / & 384.22 \\
    & MetaRCA & \textbf{0.66} & \textbf{0.54} & \textbf{0.88} & \textbf{0.82} & \textbf{0.90} & \textbf{0.84} & 0.90 \\
    \bottomrule
  \end{tabular}
  }%
\end{table*}

\subsection{RQ2: Generalizability and Scalability of MetaRCA}
A critical requirement for any practical RCA solution is its ability to generalize across diverse, evolving systems and to scale efficiently with increasing system complexity. We conducted two sets of analyses to evaluate these two facets of MetaRCA.

\begin{figure}[htbp]
    \centering
    \includegraphics[width=0.95\linewidth]{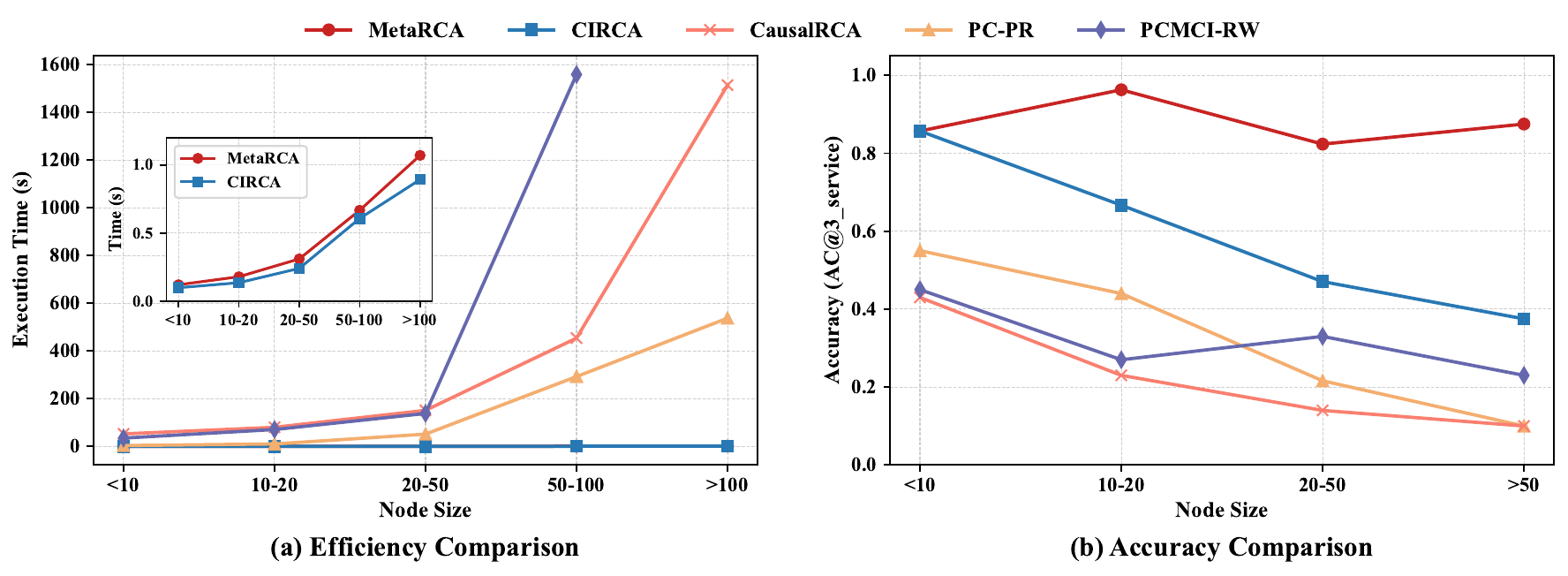}
    \caption{Performance comparison of different RCA methods as the number of microservice nodes increases. (a) Efficiency, measured by execution time. (b) Service-level localization accuracy, measured by AC@3.}
    \label{fig:onsize}
\end{figure}

\textit{Scalability.}
We first analyze the performance of all methods as the number of microservice nodes  increases. Figure~\ref{fig:onsize} presents the trends for both execution time and localization accuracy.
% \begin{figure}[htbp]
%     \centering

%     \begin{subfigure}[b]{0.48\textwidth}
%         \centering
%         \includegraphics[width=\textwidth]{samples/fig/computational_efficiency_plot.pdf}
%         \caption{Efficiency Comparison}
%         \label{fig:size_sub1}
%     \end{subfigure}
%     \hfill
%     \begin{subfigure}[b]{0.48\textwidth}
%         \centering
%         \includegraphics[width=\textwidth]{samples/fig/algorithm_accuracy_plot.pdf}
%         \caption{Accuracy Comparison}
%         \label{fig:size_sub2}
%     \end{subfigure}

%     \caption{Performance comparison of different RCA methods as the number of microservice nodes increases. (a) Efficiency, measured by execution time. (b) Service-level localization accuracy, measured by AC@3.}
%     \label{fig:onsize}
%     \Description{}
% \end{figure}

As shown in Figure~\ref{fig:onsize}(a), the efficiency of MetaRCA is a key advantage. It exhibits a sub-linear increase in execution time, which remains highly competitive and is only marginally higher than the lightweight CIRCA. This is in stark contrast to data-driven methods like CausalRCA and PC-PR, whose computational overhead grows prohibitively with system scale. This efficiency stems from our framework's localized inference process, which constrains the analysis to a small, fault-relevant subgraph rather than the entire system.

More importantly, Figure~\ref{fig:onsize}(b) shows that MetaRCA demonstrates remarkable robustness in accuracy. It maintains a high and stable AC@3 score even as the system grows to over 50 nodes. In contrast, the accuracy of all baseline methods degrades significantly with increasing scale. This suggests that their reliance on either topology alone or purely statistical discovery makes them susceptible to the increased noise and complexity of larger systems. Our framework's ability to fuse a generalizable knowledge base with real-time contextual pruning effectively filters out this noise, ensuring stable performance regardless of scale.

% \begin{figure}[htbp]
%     \centering

%     \begin{subfigure}[b]{0.48\textwidth}
%         \centering
%         \includegraphics[width=\textwidth]{samples/fig/comparison_crosssystem_service.pdf}
%         \caption{Service-Level Accuracy Distribution}
%         \label{fig:cross_sub1}
%     \end{subfigure}
%     \hfill
%     \begin{subfigure}[b]{0.48\textwidth}
%         \centering
%         \includegraphics[width=\textwidth]{samples/fig/comparison_crosssystem_metric.pdf}
%         \caption{Metric-Level Accuracy Distribution}
%         \label{fig:cross_sub2}
%     \end{subfigure}

%     \caption{Distribution of localization accuracy (AC@3) for different RCA methods across seven systems (three open-source and four production). The box plots show the performance distribution at (a) the service level and (b) the metric level.}
%     \label{fig:cross}
%     \Description{}
% \end{figure}

\begin{figure}[htbp]
    \centering
    \includegraphics[width=\linewidth]{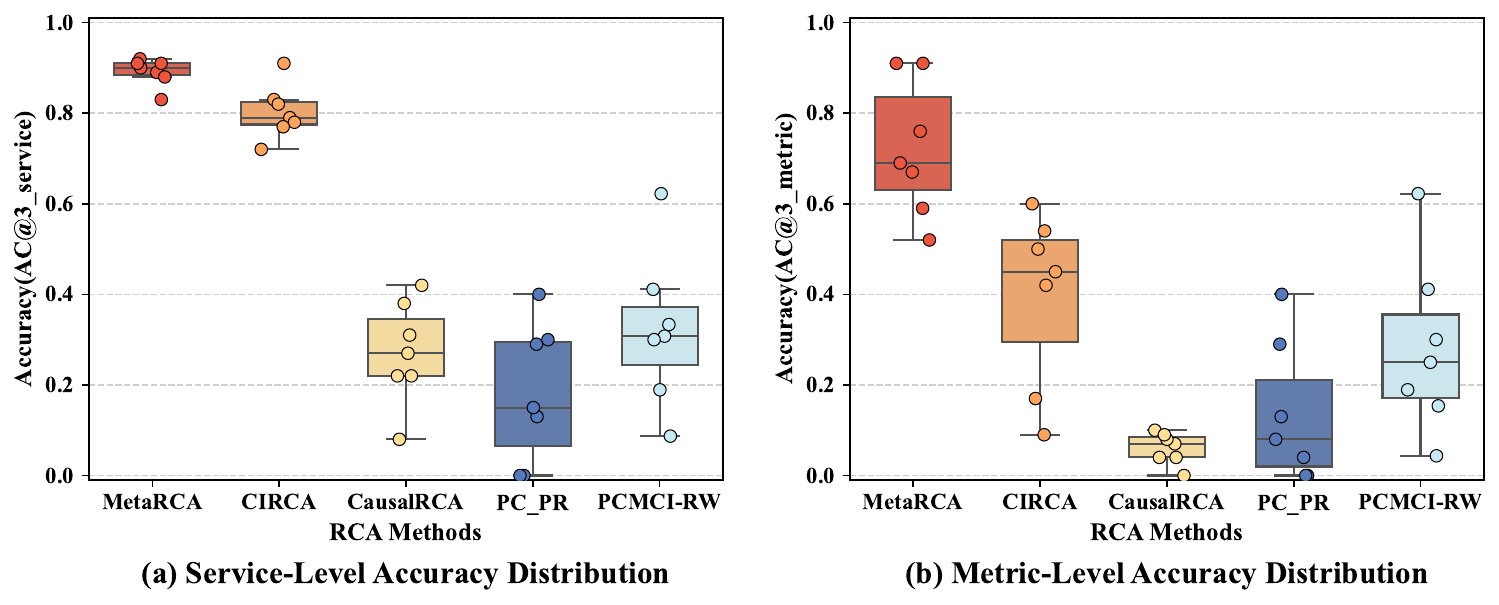}
    \caption{Distribution of localization accuracy (AC@3) for different RCA methods across seven systems (three open-source and four production). The box plots show the performance distribution at (a) the service level and (b) the metric level.}
    \label{fig:cross}
\end{figure}

\textit{Generalizability.}
To evaluate generalizability, we tested all methods across seven distinct systems (three open-source, four production). Figure~\ref{fig:cross} presents the distribution of AC@3 scores for each method.

The experimental results clearly show that while CIRCA also demonstrates robust performance, particularly at the coarse-grained service level, MetaRCA not only achieves the highest median accuracy but also exhibits a crucial advantage in the more challenging task of metric-level localization. Furthermore, the variance in MetaRCA's performance (indicated by the interquartile range of the box plot) is visibly smaller than that of the evaluated baselines, demonstrating its consistent and reliable performance across diverse system architectures. This strong generalizability is a direct benefit of our core design: the MCG captures reusable, metadata-level causal knowledge. Consequently, it can be effectively applied to new systems in a zero-shot manner, a crucial advantage over methods that require system-specific model training or are tightly coupled to a particular instance graph.

% tcolorbox is broken
% \begin{tcolorbox}

% \textbf{Answering RQ2:} \textit{MetaRCA} demonstrates excellent scalability, maintaining robust accuracy and sub-linear time overhead as system size increases. Furthermore, its knowledge-driven approach provides strong cross-system generalizability, delivering consistently high performance without the need for system-specific retraining.
% \end{tcolorbox}

\begin{center}
    \cornersize{0.15} 
    \setlength{\fboxsep}{8pt} 
    \Ovalbox{
        \begin{minipage}{0.92\columnwidth}
            \textbf{Answering RQ2:} \textit{MetaRCA} demonstrates excellent scalability, maintaining robust accuracy and sub-linear time overhead as system size increases. Furthermore, its knowledge-driven approach provides strong cross-system generalizability, delivering consistently high performance without the need for system-specific retraining.
        \end{minipage}
    }
\end{center}

\subsection{RQ3: Quality of Causal Graphs Constructed by MetaRCA}
To isolate and evaluate the quality of the causal graph generated by our approach, we designed a controlled experiment. The core idea is to fix the final ranking algorithm and vary only the input causal graph. We selected four different ranking methods for this test (PageRank\cite{howfar2024root}, Random Walk\cite{howfar2024root}, RHT(method in CIRCA\cite{li2022causal_circa}), and our previously discussed CCB(\S\ref{sec:ccb}).

\begin{figure}[htbp]
    \centering
    \includegraphics[width=0.95\linewidth]{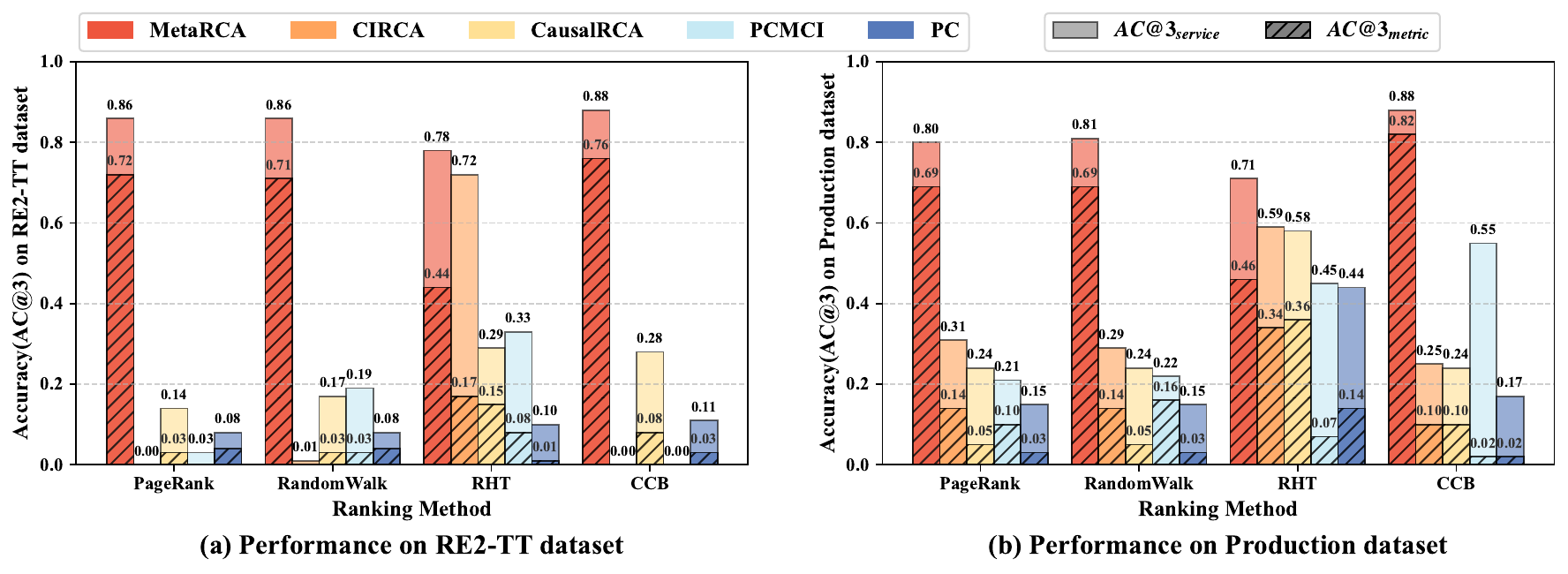}
    \caption{Localization accuracy (AC@3) of different causal graph construction and ranking method combinations. The performance is shown on (a) the RE2-TT dataset and (b) the Production dataset, with results for both service-level (total height) and metric-level (solid portion) accuracy.}
    \label{fig:diff_rank}
\end{figure}

Figure~\ref{fig:diff_rank} presents the AC@3 results on both the RE2-TT and the Production datasets. The results consistently and unequivocally demonstrate the superiority of the causal graph constructed by MetaRCA. Across all four ranking methods and on both datasets, using MetaRCA's causal graph as input yields significantly higher accuracy than using graphs generated by any of the other construction methods (CIRCA\cite{li2022causal_circa}, CausalRCA\cite{xin2023causalrca}, PC\cite{spirtes2000causation}, PCMCI\cite{pcmci_2019}).

For instance, on the \textit{Production} dataset with the PageRank ranker, MetaRCA achieves a service-level accuracy of 0.82 and a metric-level accuracy of 0.71. The next best construction method, CIRCA, only achieves 0.31 and 0.14, respectively. 

We attribute this superiority to our framework's unique ability to fuse high-level, generalizable knowledge from the MCG with real-time, context-specific evidence from observability data. This process generates a high-signal, low-noise localized graph that effectively guides the ranking algorithm, a feat that purely data-driven or topology-based methods struggle to achieve.

% \begin{tcolorbox}
% \textbf{Answering RQ3:} The causal graph generated by MetaRCA's context-aware framework, which fuses meta-causal knowledge and real-time data, is demonstrably superior to those constructed by existing methods. This high-quality graph is the primary driver of our framework's strong performance demonstrated in our evaluation.
% \end{tcolorbox}

\begin{center}
    \cornersize{0.15} 
    \setlength{\fboxsep}{8pt} 
    \Ovalbox{
        \begin{minipage}{0.92\columnwidth}
            \textbf{Answering RQ3:} The causal graph generated by MetaRCA's context-aware framework, which fuses meta-causal knowledge and real-time data, is demonstrably superior to those constructed by existing methods. This high-quality graph is the primary driver of our framework's strong performance demonstrated in our evaluation.
        \end{minipage}
    }
\end{center}

% \begin{figure}[t]
%     \centering

%     \begin{subfigure}[b]{0.48\textwidth}
%         \centering
%         \includegraphics[width=\textwidth]{samples/fig/diff_rank_TT.pdf}
%         \caption{Performance on RE2-TT Dataset}
%         \label{fig:diff_rank_sub1}
%     \end{subfigure}
%     \hfill
%     \begin{subfigure}[b]{0.48\textwidth}
%         \centering
%         \includegraphics[width=\textwidth]{samples/fig/diff_rank_industy.pdf}
%         \caption{Performance on Production Dataset}
%         \label{fig:diff_rank_sub2}
%     \end{subfigure}

%     \caption{Localization accuracy (AC@3) of different causal graph construction and ranking method combinations. The performance is shown on (a) the RE2-TT dataset and (b) the Production dataset, with results for both service-level (total height) and metric-level (solid portion) accuracy.}
%     \label{fig:diff_rank}
%     \Description{}
% \end{figure}

\subsection{RQ4: Ablation Study of Key Designs}

To quantify the individual contribution of each key design component within the MetaRCA framework, we conducted a comprehensive ablation study. We compare our full model against four variants, each with a specific component disabled. The results are presented in Table~\ref{tab:ablation_study}.

\begin{table}[h]
  \caption{Ablation study results (AC@3) on the RE2-TT and Production datasets.}
  \label{tab:ablation_study}
  \begin{tabular}{@{}lcccc@{}}
    \toprule
    \multirow{2}{*}{\textbf{Method Variant}} & \multicolumn{2}{c}{\textbf{RE2-TT}} & \multicolumn{2}{c}{\textbf{Production}} \\
    \cmidrule(lr){2-3} \cmidrule(lr){4-5}
    & service & metric & service & metric \\
    \midrule
    \textbf{MetaRCA (Full)} & \textbf{0.88} & \textbf{0.76} & \textbf{0.88} & \textbf{0.82} \\
    \midrule
    MetaRCA w/o report evidence & 0.83 & 0.63 & 0.80 & 0.72 \\
    MetaRCA w/o data evidence & 0.81 & 0.70 & 0.86 & 0.76 \\
    MetaRCA w/o any evidence & 0.80 & 0.62 & 0.79 & 0.70 \\
    MetaRCA w/o online pruning & 0.78 & 0.51 & 0.63 & 0.56 \\
    \bottomrule
  \end{tabular}
\end{table}

The results provide two key insights. First, the multi-source evidence fusion is crucial for achieving high accuracy. On the \textit{Production} dataset, incorporating evidence significantly boosts the metric-level AC@3 from 0.70(\textit{w/o any evidence}) to 0.82. Interestingly, on this dataset, the removal of report evidence (metric AC@3 drops to 0.72) causes a larger performance decrease than removing data evidence (a drop to 0.76). This suggests that for real-world failures, the knowledge extracted from historical incident reports contributes more significantly to the final accuracy.

Second, online weighting and pruning are indispensable for precise localization. Disabling this feature results in the largest performance drop. On the \textit{Production} dataset, service-level AC@3 decreases sharply from 0.88 to 0.63, and metric-level AC@3 from 0.82 to 0.56. This clearly demonstrates that even a high-quality knowledge base must be contextualized with real-time data to effectively identify the root cause. Filtering irrelevant paths allows the analysis to focus on the most plausible failure propagation chains.

% \begin{tcolorbox}
% \textbf{Answering RQ4:} Our ablation study confirms the critical contribution of each key design in MetaRCA. The fusion of multi-source evidence is essential for building a high-quality knowledge base, while the online weighting and pruning  mechanism is indispensable for translating that knowledge into accurate, context-aware root cause analysis.
% \end{tcolorbox}

\begin{center}
    \cornersize{0.15} 
    \setlength{\fboxsep}{8pt} 
    \Ovalbox{
        \begin{minipage}{0.92\columnwidth}
            \textbf{Answering RQ4:} Our ablation study confirms the critical contribution of each key design in MetaRCA. The fusion of multi-source evidence is essential for building a high-quality knowledge base, while the online weighting and pruning  mechanism is indispensable for translating that knowledge into accurate, context-aware root cause analysis.
        \end{minipage}
    }
\end{center}

\subsection{Case Study}
\label{sec:case}
\begin{figure*}[]
    \centering
    \begin{subfigure}[b]{0.24\textwidth}
        \centering
        \includegraphics[width=\linewidth]{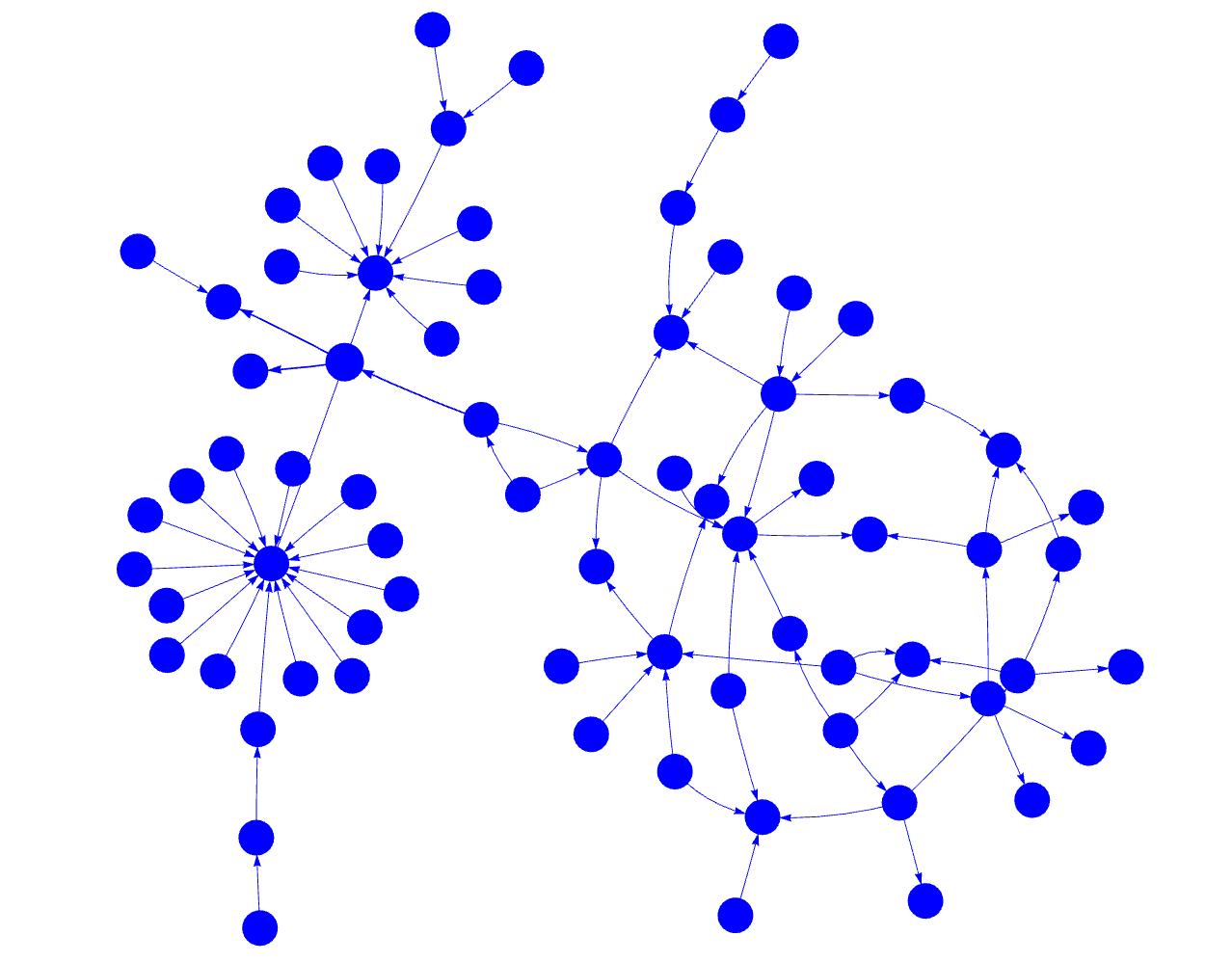}
        \caption{Complex topology of the 73-service system.}
        \label{fig:case_study_a}
    \end{subfigure}
    \hfill % 
    \begin{subfigure}[b]{0.24\textwidth}
        \centering
        \includegraphics[width=\linewidth]{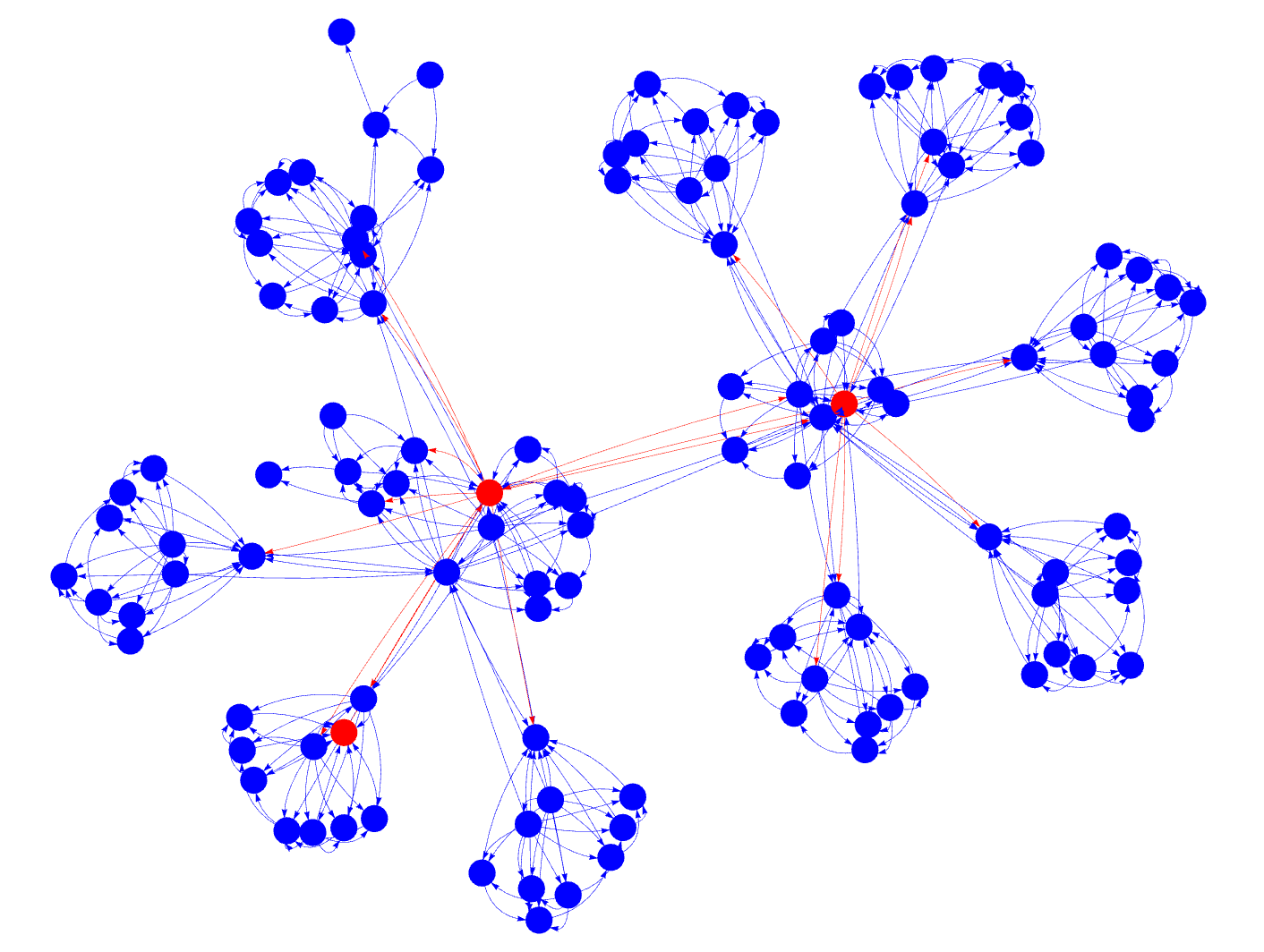}
        \caption{The noisy, naively instantiated causal graph}
        \label{fig:case_study_b}
    \end{subfigure}
    \hfill % 
    \begin{subfigure}[b]{0.24\textwidth}
        \centering
        \includegraphics[width=\linewidth]{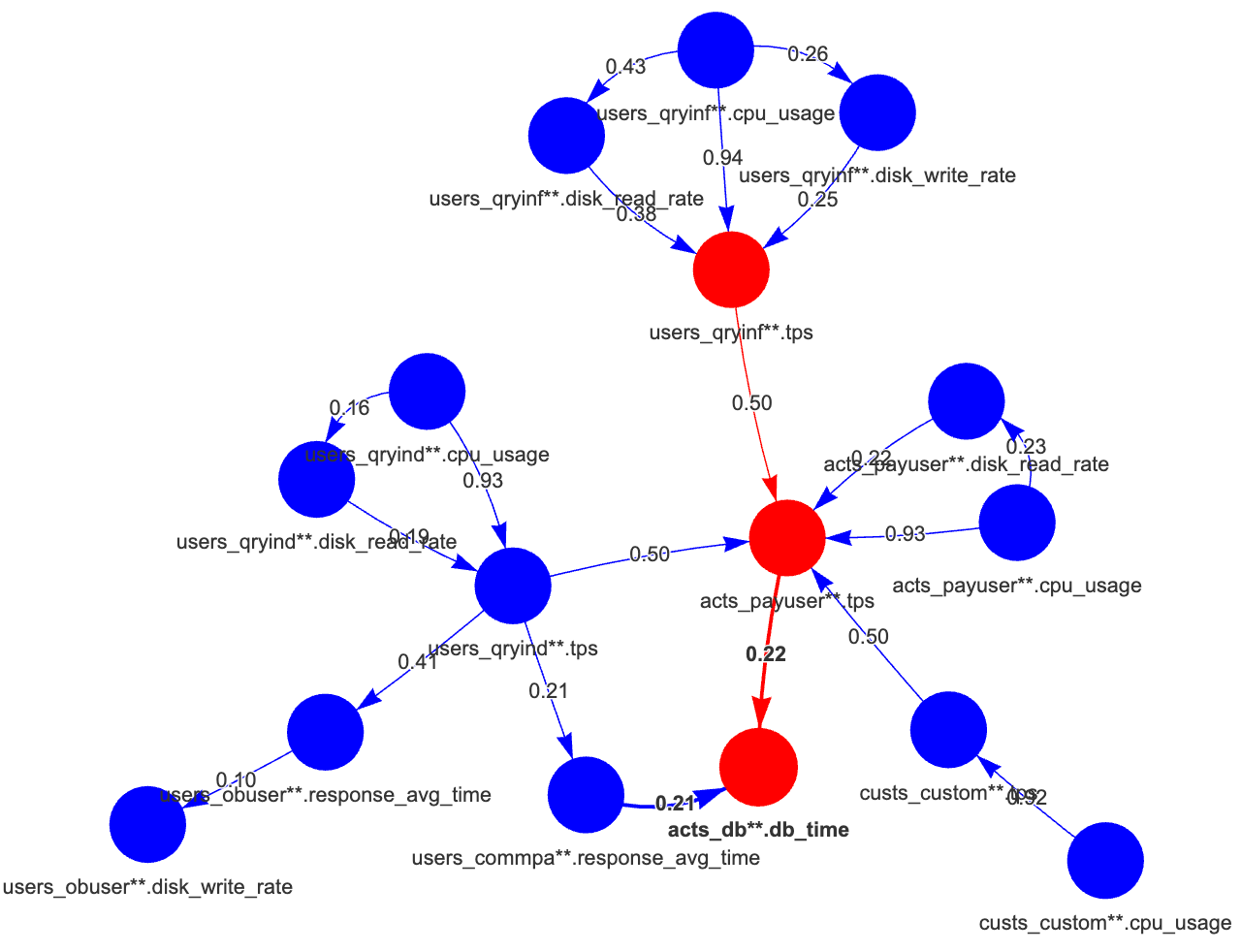}
        \caption{Causal graph after data fusion.}
        \label{fig:case_study_c}
    \end{subfigure}
    \hfill % 
    \begin{subfigure}[b]{0.24\textwidth}
        \centering
        \includegraphics[width=\linewidth]{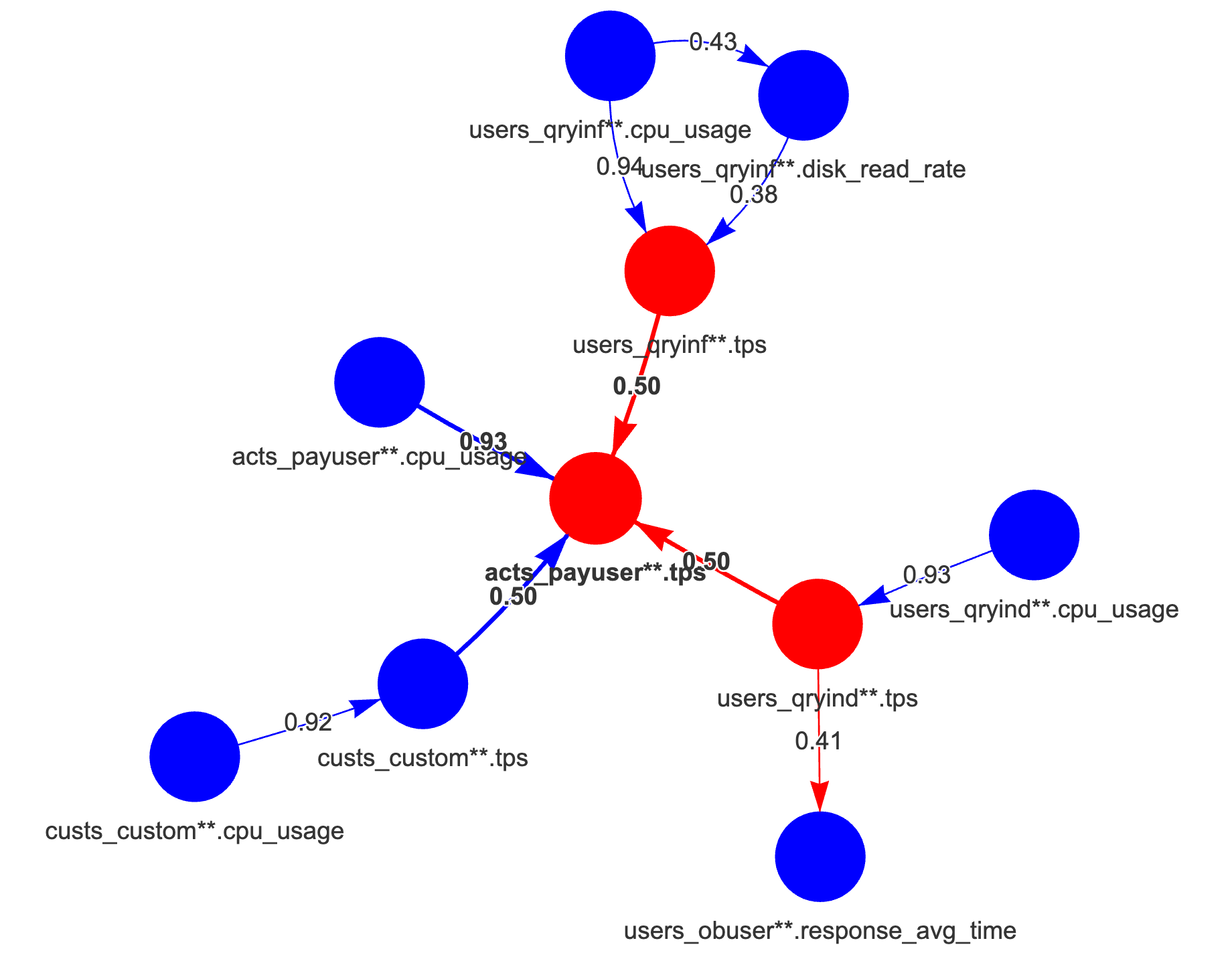}
        \caption{Pruned root cause graph.}
        \label{fig:case_study_d}
    \end{subfigure}
    
    \caption{The MetaRCA process in a real-world case.}
    \label{fig:case_study}
\end{figure*}

In an incident occurred in August 2025, triggered by an unscheduled batch operation, the  root cause was a TPS (transactions per second) surge on the ingress service, \textit{acts\_payuser**}.  The system comprises 73 services with a complex call topology, as shown in Figure~\ref{fig:case_study}(a). If the original knowledge from the MCG were naively projected onto the system, it would result in the complex initial causal graph shown in Figure~\ref{fig:case_study}(b).

Our MetaRCA method first defines the \textit{Fault Relevance Zone} using anomaly detection, then instantiates the MCG into a localized causal graph. After fusing this knowledge with real-time data, the resulting graph is shown in Figure~\ref{fig:case_study}(c) (edges with weights < 0.1 have been removed for clarity). Within this graph, it can be seen that abnormal metrics from two services point to db\_time of \textit{acts\_db**}. Many diagnostic methods would misidentify the low-weight db\_time metric as the root cause, despite its increase being only a minor, cascading effect.

Crucially, MetaRCA's data fusion and pruning mechanism correctly identifies and removes this interfering edge. This results in a cleaner, more symptom-focused causal graph, as shown in Figure~\ref{fig:case_study}(d), enabling the successful localization of the true root cause. This case study demonstrates that in large-scale, complex production systems, an initial fault can trigger numerous chain reactions and misleading correlations. Our framework's multi-stage process of instantiation and data-driven refinement is key to managing this complexity.

\section{Discussion}
\label{sec:discussion}
\textbf{Discussion on hyperparameters.}
Our framework incorporates three key hyperparameters: the temporal decay constant $k$, the base impact strengths $\lambda$ and online pruning threshold $\theta_p$. The temporal decay mechanism ($k=0.005$) is a deliberate design choice to address concept drift in evolving systems. It prevents the saturation of belief scores from the sheer accumulation of historical evidence and prioritizes recent, more relevant information. The value is tied to the temporal unit (days) and can be tuned based on a system's evolutionary pace. The asymmetric weighting ($\lambda_{fr}=0.5 \gg \lambda_{da}=0.05$) reflects the intrinsic fidelity of the evidence sources. Case evidence from expert-validated incident reports represents a high-fidelity, low-noise signal, while statistical evidence from automated discovery is inherently of lower fidelity and higher noise. While the exact values were tuned on a separate validation set, the core principle is that the relative importance of these evidence types is maintained.
Finally, the online pruning threshold, $\theta_p$, A threshold set too low results in a noisy graph that diminishes the ranking algorithm's effectiveness, while one set too high risks erroneously pruning the true causal path. Our empirical analysis indicated that a value in the range of [0.2, 0.4] offers the best balance, leading to our choice of $\theta_p = 0.3$ for the final evaluation. 

\textbf{Threats to validity.}
We acknowledge several potential threats to our study's validity. (1) Our conclusions depend on the correct implementation and fair comparison of baselines. We mitigated this by meticulously reproducing all methods from their original research papers and applying consistent evaluation metrics.  (2) While we validated our MCG on systems with similar technology stacks, its effectiveness on different architectures (e.g., big data frameworks) and its sensitivity to the quality of input knowledge require further investigation. (2) Our primary evaluation metrics, \textit{AC@k} and \textit{Average RCA Time}, are standard but serve as proxies for the ultimate goal of reducing mean time to repair (MTTR).

\section{Related Work} 
Our work intersects with two primary research areas: causality-based RCA, knowledge acquisition for causal graphs.

\textbf{Causality-based Root Cause Analysis.}
The use of causal graphs for RCA has evolved along several lines. A foundational approach involves purely data-driven causal discovery from monitoring metrics using algorithms like PC \cite{spirtes2000causation}, Granger \cite{granger1980testing}, or its variants \cite{runge2019detecting}, as seen in systems like \cite{wang2018cloudranger, meng2020localizing}. However, these methods often face scalability issues and can infer spurious relationships without structural priors \cite{howfar2024root, li2022causal_circa, liu2021microhecl,Microscope}. A dominant trend emerged that constrains the search space using the service call graph, effectively integrating system topology \cite{chen2014causeinfer, li2022causal_circa}. 
%For example, \cite{xin2023causalrca} employs a gradient-based method for learning the causal structure and quantifying the causal effect weights in a graph. 
A parallel paradigm, exemplified by Groot \cite{wang2021groot} and CoE \cite{yao2024chain}, models system behavior into an event-based causal graph. However, the causal graphs in these methods are typically instance-level and cannot achieve strong generalization while leveraging historical data. This is the core problem that MetaRCA addresses by abstracting knowledge to the metadata level.

\textbf{Knowledge Acquisition for Causal Graphs.}
Purely data-driven causal discovery often struggles with scalability\cite{howfar2024root} and is highly sensitive to data quality\cite{zhang2024survey}. Integrating various forms of knowledge has become a key strategy for enhancing the fidelity of causal graphs\cite{wang2024kgroot,li2022causal_circa}. 
One approach is to leverage high-quality domain expertise, whether encoded as explicit rules and event schemas \cite{wang2021groot,yao2024chain} or as implicit structural assumptions\cite{li2022causal_circa,liu2021microhecl} based on topology. While accurate, this knowledge is often manual and brittle. 
To automate acquisition, another line of research applies NLP to mine technical artifacts like incident reports \cite{saha2022mining,causalitymining}.More recently, Large Language Models have been used to bootstrap causal discovery\cite{kiciman2023causal,xie2024cloudatlasefficientfault} or or directly employed for RCA tasks\cite{chen2024automatic, xu2025openrca}, but their practical use is challenged by factual "hallucination" and high inference costs. MetaRCA contributes by proposing a systematic framework that fuses the outputs of LLM bootstrapping, validated case evidence from reports, and statistical evidence from data, thereby creating a more robust and reliable knowledge base.

\section{Conclusion}
In this study, we introduce MetaRCA, a generalizable RCA framework that addresses the critical challenges of calability, generalization and knowledge integration in cloud-native systems. MetaRCA first constructs a Meta Causal Graph offline, for which we design an evidence-driven belief evolution algorithm that fuses knowledge from LLMs, incident reports, and statistical analysis. For online diagnosis, dynamically instantiates the MCG into a localized graph that is weighted and pruned  with real-time data, before root cause ranking. We conducted a comprehensive evaluation of MetaRCA on both public benchmarks and real-world production dataset, assessing its performance, scalability and generalizability. The experimental results demonstrate that MetaRCA achieves superior localization accuracy and efficiency compared to baseline methods, an advantage that becomes more pronounced as system scale increases. Furthermore, our cross-system evaluation confirms that MetaRCA exhibits robust generalizability, delivering stable and adaptive performance across diverse architectures without requiring system-specific retraining.

\begin{acks}
This work was supported in part by
National Key Research and Development Program of China
(Grant Number: 2024YFB4505904), the National Natural Science Foundation of China under Grant 62272495 and the
Guangdong Basic and Applied Basic Research Foundation
under Grant 2023B1515020054. The corresponding author is Pengfei Chen.
\end{acks}

\appendix
\section{Prompt Design}
\label{sec:prompt}

\begin{figure}[H]
    \centering
    \includegraphics[width=0.99\linewidth]{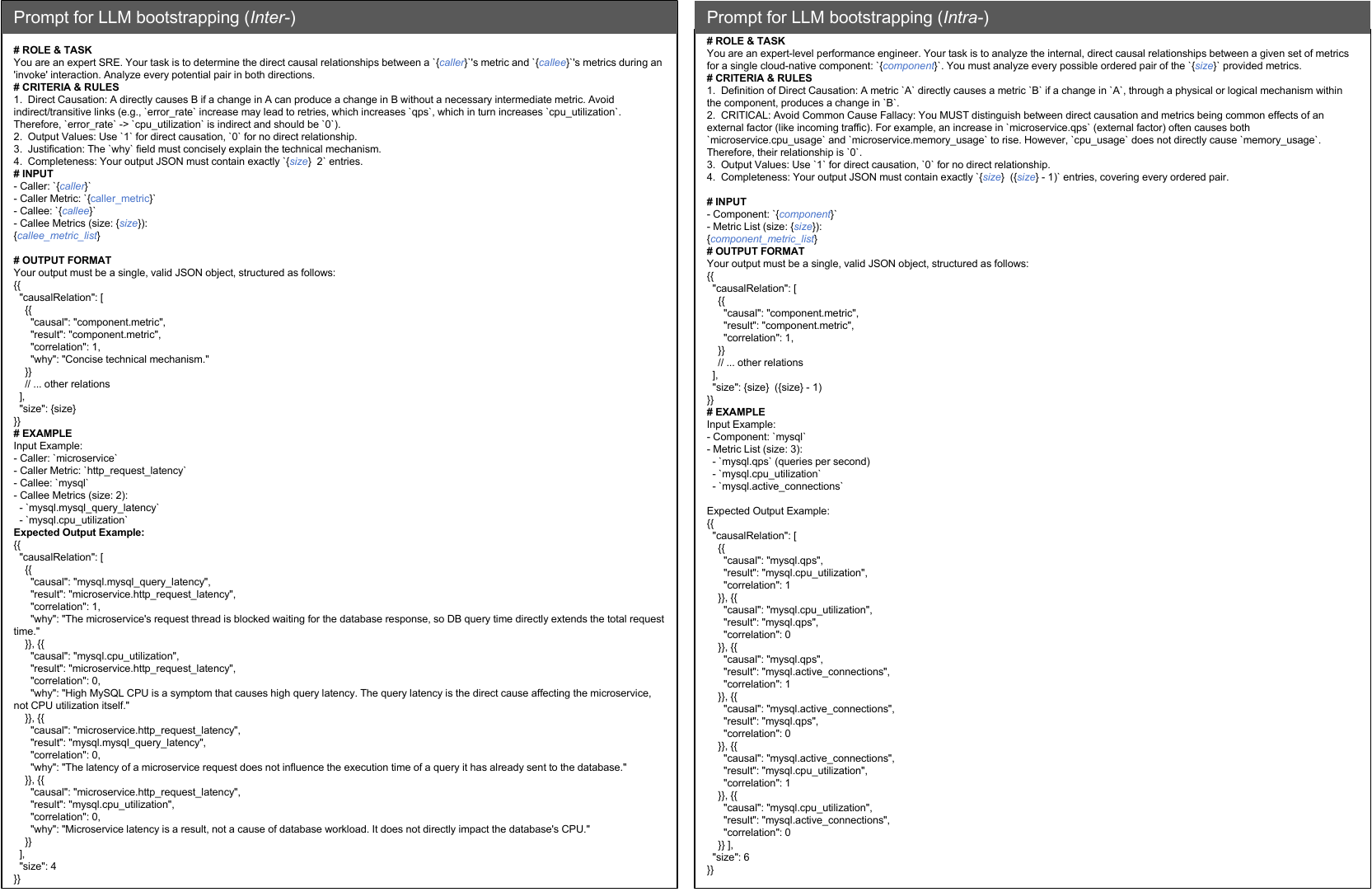}
    \caption{Prompt template for LLM bootstrapping }
    \label{fig:prompt_boot}
\end{figure}

\begin{figure}[H]
    \centering
    \includegraphics[width=0.99\linewidth]{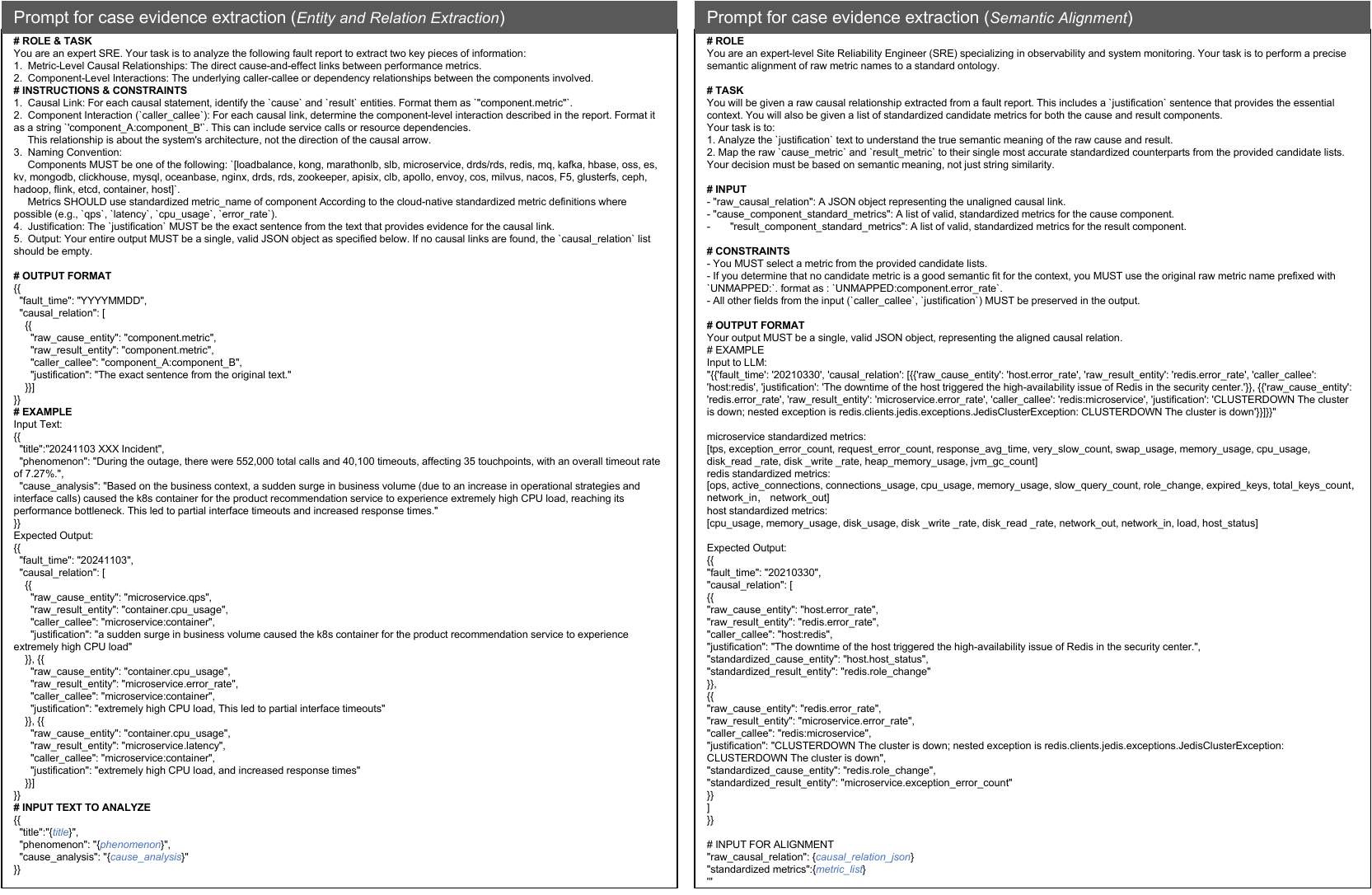}
    \caption{Prompt templates for evidence extraction }
    \label{fig:prompt_extract}
\end{figure}

Figures \ref{fig:prompt_boot} and \ref{fig:prompt_extract} illustrate the prompt templates for the LLM bootstrapping and evidence extraction tasks, respectively.

\bibliographystyle{ACM-Reference-Format}
\bibliography{sample-base}

\end{document}